\documentclass[12pt]{article}

\usepackage{amsmath}
\usepackage{graphicx,psfrag,epsf}
\usepackage{enumerate}
\usepackage{natbib}
\usepackage{url} % not crucial - just used below for the URL

\newcommand{\blind}{1}

\usepackage[titletoc,title]{appendix}
\usepackage{authblk}
\usepackage{rotating, lineno, graphicx, color, setspace, amssymb, amsmath, amsthm, amstext, booktabs, fancyhdr, multirow, float, bm, xr, color}
\usepackage{geometry}

\renewcommand{\baselinestretch}{1.5}

\newtheorem{theorem}{Theorem}

\newtheorem{corollary}{Corollary}

\newtheorem{remark}{Remark}
\newtheorem{condition}{Condition}

\DeclareMathOperator{\sgn}{sgn}

\newcommand{\trans}{\mathrm{T}}

\newcommand{\bM}{\mathbf{M}}

\begin{document}

\def\spacingset#1{\renewcommand{\baselinestretch}%
{#1}\small\normalsize} \spacingset{1}

%%%%%%%%%%%%%%%%%%%%%%%%%%%%%%%%%%%%%%%%%%%%%%%%%%%%%%%%%%%%%%%%%%%%%%%%%%%%%%

\if1\blind
{
  \title{\bf Multi-threshold Change Plane Model: Estimation Theory and Applications in Subgroup Identification}
  \author{Jialiang Li\\
    Department of Statistics and Applied Probability, National University of Singapore\\
  %\thanks
  %{ % The authors gratefully acknowledge \textit{please remember to list all relevant funding sources in the unblinded version}}\hspace{.2cm}\\
    %and \\
   Yaguang Li and  Baisuo Jin \\
   Department of Statistics and Finance, University of Science and Technology of China}
   \date{}
  \maketitle
} \fi

\if0\blind
{
  \bigskip
  \bigskip
  \bigskip
  \begin{center}
    {\LARGE\bf Multi-threshold Change Plane Model: Estimation Theory and Applications in Subgroup Identification}
\end{center}
  \medskip
} \fi

\bigskip
\begin{abstract}
We propose a multi-threshold change plane regression model which naturally partitions the observed subjects into subgroups with different covariate effects. The underlying grouping variable is a linear function of covariates and thus multiple thresholds form parallel change planes in the covariate space.  We contribute a novel 2-stage approach to estimate the number of subgroups, the location of thresholds and all other regression parameters. In the first stage we adopt a group selection principle to consistently identify the number of subgroups, while in the second stage change point locations and model parameter estimates are refined by a penalized induced smoothing technique. Our procedure allows sparse solutions for relatively moderate- or high-dimensional covariates. We further establish the asymptotic properties of our proposed estimators under appropriate technical conditions.  We evaluate the performance of the proposed methods by simulation studies and provide illustration using two medical data. Our proposal for subgroup identification may lead to an immediate application in personalized medicine.
\end{abstract}

\noindent%
\vfill

\newpage
\spacingset{1.45} % DON'T change the spacing!

\section{Introduction}

Individualized learning and modeling has become increasingly important in statistics and computer science, especially for solving the personalized medical
treatment problems.  The traditional  ``one size fits all" approach is unable to detect important patterns in the sub-populations and make the best personalized
predictions for specific individuals. For example, in the fight against cancer and other diseases, it is difficult to recommend a treatment that works for all
patients. Consequently the rise of precision medicine and analysis of electronic health record data motivates researchers to identify meaningful subgroups and
model the relationships between response and predictors differently across the subgroups.

Earlier development in personalized medicine focused on determining dynamic treatment regimes at multiple stages. Popular model based methods for estimating the
optimal individualized treatment regimes include the Q-learning \citep{qian2011performance, goldberg2012q} and the A-learning \citep{murphy2003optimal,
robins2004optimal, schulte2014q}, which models interactions between the treatments and covariates and is more robust to model misspecification than Q-learning.
\cite{zhao2012estimating} introduced the framework of outcome weighted learning (O-learning) to directly find the optimal binary treatment rule from a
classification perspective.  Other relevant works include \cite{zhang2012robust}, \cite{zhang2013robust} and \cite{zhao2015new} among many others. Recently,
\cite{wager2017estimation} developed a forest-based method for treatment effect estimation, \cite{Fan2017concordance} proposed a concordance-assisted learning
and \cite{Jiang2017treatment} was via maximize survival probability to estimate optimal treatment regimes.

In addition to these optimization-involved learning strategy, another burgeoning research direction in personalized medicine is categorizing patients into
subgroups using appropriate algorithms and then consider the treatment effects for those subgroups. Many data-driven approaches for subgroup identification have
been developed in the literature. One commonly used approach is the tree-based method. Early works include Automatic Interaction Detection (AID) \citep{MS1963}
and theta automatic interaction detection (THAID) \citep{MM1972}. \cite{L2002} developed the generalized unbiased interaction detection and estimation (GUIDE)
method to identify subgroups of subjects for whom the treatment has an enhanced effect. \cite{FJR2011} proposed a virtual twins (VT) method to obtain the
subgroups with an enhanced treatment effect.  \cite{cai2011analysis} and  \cite{zhao2013effectively} used a parametric scoring system to estimate
subject-specific treatment differences and then identify a promising population who benefit more from the new treatment. \cite{shen2015inference} adapted a
finite logistic-normal mixture model to subgroup analysis by a likelihood-based test. \cite{chen2017general} propose a general framework for subgroup
identification by weighting and A-learning approaches. In fact, all the aforementioned works used similar techniques to those in change point analysis
\citep{bai1997estimation} and can be justified rigorously using the traditional change point theory. Recently, \cite{fan2017change} considered a change plane
method to test the existence of subgroup using a doubly robust score statistic. The advantage of change plane over change point is that we may allow the
underlying grouping variable to be a linear combination of covariates in stead of a single covariate. However, the approach in \cite{fan2017change} only allows
a single threshold (and thus only two subgroups) and searching the supremum of squared score test statistics over a unit ball could be quite challenging,
especially when aiming for multiple groups.

%In addition, as more information per individual is being collected in clinical studies and not all of the information is relevant for subgroup detection.

To formally address the issue in this paper, we will consider a change plane model with unknown number of thresholds, which extends the familiar change point
threshold regression model. In fact the change point model or the so-called segment regression has wide applications in economics \citep{tong1990non,
li2012least,kourtellos2016structural} where the underlying grouping variable is usually the time point or a chosen regressor and subgroups are identified as the
grouping variable moves across thresholds. For a single threshold change point model, \cite{hansen2000sample} developed the asymptotic results for the threshold
parameter estimator based on the diminishing effect assumption. \cite{SEO2007704} proposed a smoothed least squares estimator and established the consistency
and asymptotic normality following the well-known smoothed maximum score estimator (\cite{horowitz1992smoothed}). Detecting multiple thresholds is a much more
challenging problem since one needs to first figure out the number of thresholds and then determine their exact locations. Recently \cite{jin2015multi} proposed
a penalty-based framework for the accelerated failure time regression model. They formulated the threshold problem as a group model selection problem and
applied the fast computing tool in \cite{jin2013novel}. However, other than \cite{fan2017change}, there is little work on change plane analysis where the
functional form of the grouping variable needs to be constructed as well as the separating threshold.

Our model allows multiple change planes which automatically generates subgroups with different covariate effects, naturally facilitating personalized medicine
and other similar applications. The technical merits of our contribution mainly lie in the following three aspects. First, instead of using only a pre-assigned
index variable in a change point model \citep{jin2015multi}, the notion of change plane grants a linear combination of the covariates and may lead to more
meaningful definition of subgroups. This framework may offer a more flexible tool for precision medicine than earlier proposals. The inference for plane-related
parameters is not standard and requires a rather technical justification. Second, our change plane model may include multiple unknown structural changes. This
is another non-trivial improvement from single threshold models because of the difficulty in determining the number of break points. A fast splitting strategy
is developed to convert the threshold identification problem into a model selection problem. We then carry out a rigorous study to argue the consistency. Third,
we notice that in practice the subgroups may only differ in covariate effects for a few selected covariates and share the same effects for others. We thus allow
some enhance effects to be zero and aim to obtain sparse solutions\citep{lu2013variable, xu2015regularized, song2015sparse}. This is achieved through a
penalized induced smoothing estimation approach. We provide the consistency of subgroup detection and asymptotic theory for such penalized estimates.

The rest of this paper is arranged as follows. In Section \ref{sec:scpl}, a penalized induced smoothing estimation is proposed for the single threshold change
plane model. In Section \ref{sec:mcpl}, the multi-threshold change plane regression for subgroup detection is formulated. We propose an iterative two-stage
procedure to detect the change planes and estimate model parameters. The theoretical properties of our procedure are established rigorously under technical
conditions. The finite-sample performance of the estimators is investigated by simulation studies in Section \ref{sec:sim}. Two empirical applications are
presented in Section \ref{sec:real}. A discussion concludes Section \ref{sec:dis}.

Throughout the paper, $\bm{1}_{q} = (1, \ldots , 1)^\trans $ is the $q$-dimensional vector of ones, $\bm{1}(\cdot)$ is an indicator function, $I_{q}$ is the
$q\times q$ identity matrix,  and $\bm{X}^\trans $ is the transpose of a matrix $\bm{X}$. For a vector $\bm{a}$, $\bm{a}^\trans$ is its transpose, $a_{j}$ is
its $j$th component, and $|\bm{a}|$, $\|\bm{a}\|$ and $\|\bm{a}\|_{\infty}$ are respectively its $L_1$-norm, $L_2$-norm (Euclidean norm) and $L_{\infty}$ norm.
For any matrix $\bM$, $\Vert \bM\Vert_{\max} = \max\{|M_{ij}|\}$ denotes the matrix max norm. If $\mathcal{A}$ is a set, its complement and its size are
respectively denoted by $\mathcal{A}^{c}$ and $|\mathcal{A}|$. In addition,  ``$\rightarrow_{a.s}$'' denotes convergence with probability 1 and
"$\overset{D}{\to}$" denotes convergence in distribution.

%\section{Change-Plane Analysis}

\section{Single threshold change plane (SCPL)}
\label{sec:scpl}

%Let $\bm{X}$ denote the covariates collected for a subject in an experimental or observational study. Denote $\bm{Z}$ to be the variables to be used for grouping, which is usually a subset of $\bm{X}$. The observed data consist of $\lbrace (\bm{X}^{\trans}_i, \bm{Z}^{\trans}_i, Y_i)^{\trans}, i = 1, \dots, n \rbrace$, which are independent and identically distributed.

We denote ${Y}_i\in \mathbb{R}$ to be the response variable of interest for the $i$th subject in a sample of size $n$. We first consider the following single
threshold change plane model:
\begin{align} \label{scp}
	Y_{i} =& \bm{X}^{\trans}_i\bm{\beta} + \bm{X}^{\trans}_i\bm{\delta}\bm{1}(\bm{Z}^{\trans}_i\bm{\theta} \geq 0) + \epsilon_i, ~~\text{$i = 1, \dots, n$},
\end{align}
where $\bm{X}_i=(X_{i1}, \dots, X_{ip})^{\trans}$ is a $p$-dimensional vector, regression coefficients $\bm{\beta} = (\beta_1, \dots, \beta_p)^{\trans}$ and
$\bm{\delta} = (\delta_1, \dots, \delta_p)^{\trans}$ are the covariate effects for the baseline group and the effect differences between the two groups. We also
observe the grouping variables $\bm{Z}_i \in \mathbb{R}^{d+1}$ where  the first element of $\bm{Z}_i$ is assumed to be constant one. The corresponding
coefficient $\bm{\theta} = (\theta_0, \theta_1, \dots, \theta_d)^{\trans}$ is a $(d+1)$-dimensional vector. For the sake of identifiability, we assume that
$\|\bm{\theta}\| = 1$. We assume $E(\epsilon_i|\bm{X}_i) = 0$ and does not impose additional distribution assumption on the error terms.
$\bm{\delta}$ is an enhanced treatment effect with which a subgroup is defined by the change-plane $\bm{1}(\bm{Z}^{\trans}\bm{\theta} \geq 0 )$. If
$\bm{\delta} = \bm{0}$, then the parameter $\bm{\theta}$ is not identified.

Similar models have been considered in \cite{SEO2007704} and \cite{fan2017change} where only two comparison groups are assumed. This kind of model itself may be
of interest in many clinical applications and therefore we provide a new yet relatively simple solution first. In the next section we will consider more general
multi-threshold model which appeals to more sophisticated procedures.

  Denote $\bm{\gamma}=(\bm{\beta}^{\trans}, \bm{\delta}^{\trans})^{\trans}$. The unknown parameters $\bm{\gamma}$ and $\bm{\theta}$ in model (\ref{scp}) can be
  estimated by minimizing the following objective function
\begin{equation} \label{objcp}
 \frac{1}{n}\sum\limits_{i=1}^{n}\lbrack Y_i -  \bm{X}_i^{\trans}\bm{\beta} - \bm{X}_i^{\trans}\bm{\delta}\bm{1}(\bm{Z}_i^{\trans}\bm{\theta} \geq 0) \rbrack^2
\end{equation}
 with constraint $\|\bm{\theta}\| = 1$. In addition, when $p$ is large we usually assume a sparse structure for $\bm{\gamma}$. Applying a penalization approach,
 we may obtain $(\tilde{\bm{\gamma}}^{\trans}, \tilde{\bm{\theta}}^{\trans})^{\trans} = \arg\min_{\|\bm{\theta}\|=1}	Q_n(\bm{\gamma}, \bm{\theta})$, where
\begin{align}\label{pen.objcp}
	Q_n(\bm{\gamma}, \bm{\theta}) = \frac{1}{n}\sum\limits_{i=1}^{n}\lbrack Y_i -  \bm{X}_i^{\trans}\bm{\beta} -
\bm{X}_i^{\trans}\bm{\delta}\bm{1}(\bm{Z}_i^{\trans}\bm{\theta} \geq 0) \rbrack^2 + p_{\lambda_n}(|\bm{\gamma}|),
\end{align}
and $p_{\lambda_n}(\cdot)$ is a penalty function with a regularization parameter $\lambda_n>0$. For the simplicity of presentation, we only consider two
well-studied non-concave penalty functions in this paper, namely the smoothly clipped absolute deviation (SCAD, \citealt{fan2001variable}) penalty and the
minimax concave-plus penalty (MC+, \citealt{zhang2010nearly}). Other penalty functions such as Lasso may also be employed.

%\subsection{The Penalized Induced Smoothing Estimation}

Directly minimizing (\ref{objcp}) or (\ref{pen.objcp}) is possible via quadratic programming but such a numerical solution may be time-consuming and highly
variable. We consider an iterative estimation procedure which may yield relatively stable solutions. For a given ${\bm{\theta}}$, model (\ref{scp}) can be
simply treated as a piecewise linear model, then the baseline coefficients $\bm{\beta}$ and enhanced effects $\bm{\delta}$ can be estimated by the penalized
least squares. On the other hand, given ${\bm{\gamma}}$, the objective function (\ref{objcp}) or (\ref{pen.objcp}) is not continuous and finding its minimizer
is still difficult. One way to overcome this difficulty is to approximate the discontinuous objective function with a smooth function
(\citealt{johnson2009induced, SEO2007704}). We can show that the estimated results of the smoothed objective function have very similar asymptotic properties as
those from the original non-smooth version.

Denote $\Phi(\cdot)$ to be the distribution function of the standard normal variable. We use $\Phi(\cdot/h)$ as a smooth approximation to the indicator
function, where the bandwidth $h$ is chosen to converge to zero as the sample size $n$ increases. Note that if $\bm{Z}^{\trans}_i\bm{\theta} >0 $,
$\Phi((\bm{Z}^{\trans}_i \bm{\theta})/h)\rightarrow 1 $ as $h\rightarrow 0$. Thus we may consider the following approximate penalized objective function for the
estimation
\begin{equation} \label{smobj}
	Q^{*}_n(\bm{\gamma}, \bm{\theta}) = \frac{1}{n}\sum\limits_{i=1}^{n}\lbrack	Y_i - \bm{X}^{\trans}_i\bm{\beta}
-\bm{X}^{\trans}_i\bm{\delta}\Phi((\bm{Z}^{\trans}_i\bm{\theta})/h) \rbrack^2 +  p_{\lambda_n}(|\bm{\gamma}|).	
\end{equation}
 This becomes a relatively standard nonlinear least squares problem (\citealt{golub2003separable}). Denote $(\tilde{\bm{\gamma}}^{*\trans},
 \tilde{\bm{\theta}}^{*\trans})^{\trans} = \arg\min_{\|\bm{\theta}\|=1} {Q}^{*}_n(\bm{\gamma}, \bm{\theta})$. The smoothed objective function, $Q^*(\bm{\gamma},
 \bm{\theta})$, is now continuously differentiable and standard numerical methods such as the Newton-Raphson algorithm can be used to efficiently compute
 $\tilde{\bm{\theta}}^{*}$. Our estimation procedure is described in details as follows.

\begin{itemize}
    \item {\sl Step 0}: Given an initial estimate of $\bm{\theta}$, say $\tilde{\bm{\theta}}_{int}^{*}$, and set $\tilde{\bm{\theta}}^{*[0]} =
        \tilde{\bm{\theta}}_{int}^{*}/\|\tilde{\bm{\theta}}_{int}^{*}\|$. Then, obtain $\tilde{\bm{\gamma}}^{*[0]}$by ordinary least squares.
  	\item {\sl Step 1}: Given $\tilde{\bm{\gamma}}^{*[k]} = (\tilde{\bm{\beta}}^{*[k]\trans}, \tilde{\bm{\delta}}^{*[k]\trans})^{\trans}$, estimate
  $\bm{\theta}$ by solving
$$\tilde{\bm{\theta}}^{*[k+1]} = \arg\min\limits_{\|\bm{\theta}\| = 1} \left\lbrace	\frac{1}{n}\sum\limits_{i=1}^{n}\lbrack	Y_i -
\bm{X}^{\trans}_i\tilde{\bm{\beta}}^{*[k]} -\bm{X}^{\trans}_i\tilde{\bm{\delta}}^{*[k]}\Phi((\bm{Z}^{\trans}_i\bm{\theta})/h) \rbrack^2 )\right\rbrace.$$

    \item {\sl Step 2}: Given $\tilde{\bm{\theta}}^{*[k+1]}$ ,  write $\mathbb{X}^{[k+1]}_{i} =
        (\bm{X}^{\trans}_{i},\bm{X}_{i}^{\trans}\Phi(\bm{Z}_i^{\trans}\tilde{\bm{\theta}}^{*[k+1]}/h))^{\trans}$, estimate $\bm{\gamma}$ by minimizing the
        following regularized least squares with a SCAD or MC+ penalty
    $$\tilde{\bm{\gamma}}^{*[k+1]} = \arg\min \left\lbrace	\frac{1}{n}\sum\limits_{i=1}^{n}\lbrack	Y_i - \mathbb{X}^{[k+1]}_{i}\bm{\gamma}\rbrack^2 +
    p_{\lambda_n}(|\bm{\gamma}|)\right\rbrace .$$
    \item {\sl Step 3}: Repeat Step 1 and Step 2 until convergence.
\end{itemize}

\begin{remark}
In Step $1$, a modified Newton-Raphson algorithm can be used to estimate $\tilde{\bm{\theta}}$ by normalizing $\bm{\theta}$ in every iteration. In practice, one
can adopt function {\tt BBoptim} in R package {\sf BB} to optimizing a high-dimensional nonlinear objective function. More detailed descriptions of separable
nonlinear least squares problems and the convergence properties of related algorithms can be found in \cite{golub2003separable} and references therein. In Step
2, $\tilde{\bm{\gamma}}^{*[k+1]}$ can be obtained by the efficient coordinate descent algorithms (\citealt{breheny2011coordinate}). Moreover, other penalty
methods can also be applied, such as the weighted lasso (\citealt{lee2016lasso}). The tuning parameters $\lambda_n$ can be chosen by the Bayesian information
criterion (BIC) or generalized cross validation (GCV). We use BIC in the numerical studies of this paper.
\end{remark}

\section{Multi-threshold change planes (MCPL)}
\label{sec:mcpl}

\subsection{Model and estimation}\label{subsec:split}
With a slight abuse of notation, we use ${\bm Z}_i$ in the following presentation to denote the $d-$vector of grouping variables without the intercept one. We
now consider change plane model with multiple thresholds and assume $\lbrace (Y_i, \bm{X}^{\trans}_i, \bm{Z}^{\trans}_i)^{\trans}$, $i = 1,\dots , n \rbrace$
follows the change-plane model with $s$ thresholds located at $a_1 < a_2 < \dots < a_s$:
\begin{align} \label{mcp}
	Y_{i} =& \bm{X}^{\trans}_i \left\lbrack \bm{\beta} + \sum_{j=1}^{s} \bm{\delta}_{j}\bm{1}(a_j < \bm{Z}^{\trans}_i\bm{\theta} \leq a_
	{s+1})\right\rbrack + \epsilon_i, ~~\text{$i = 1, \dots, n$},
\end{align}
where $\bm{\theta}$ is the change-plane parameter, $\bm{\beta}$ is the vector of coefficients for the baseline group and $\bm{\delta}_j$ is the vector of
enhanced effects for the $j$th subgroup relative to the baseline group. In this case $s \geq 0$ is also unknown and needs to be estimated and $a_1 , \dots ,
a_s$ are the threshold locations. We set $a_0 = - \infty $, and $a_{s+1} = \infty$. $\epsilon_i$'s are independent random errors with mean zero and variance
$\sigma^2$. To identify the model, we need to assume $ \bm{\theta} \in \bm{\Theta} = \{ \bm{\theta} \in \mathbb{R}^{d}: \|\bm{\theta}\| = 1, \theta_r>0, 1\le r
\le d\}$ with the $r$-th element being positive.

Denote $\bm{\eta}=(\bm{\beta}^{\trans}, \bm{\delta}^{\trans}, \bm{a}^{\trans}, \bm{\theta}^{\trans})^{\trans}$. If $a_j, j = 1,\ldots , s$ are known, then  the
unknown parameters $\bm{\beta}, \bm{\delta}$, and $\bm{\theta}$ can be estimated by minimizing the following least squares objective function with constraint
$\|\bm{\theta}\| \in{\bm\Theta}$,
\begin{equation} \label{objmcp}
	L_n(\bm{\eta}) = \frac{1}{n}\sum\limits_{i=1}^{n} \left\lbrack Y_i -  \bm{X}^{\trans}_i\bm{\beta} - \bm{X}^{\trans}_i\sum_{j=1}^{s}
\bm{\delta}_{j}\bm{1}(a_j < \bm{Z}^{\trans}_i\bm{\theta} \leq a_{s+1}) \right\rbrack^2.
\end{equation}
In general, however, the number of change-planes $s$ and the locations are all unknown. Estimation and establishing the relevant limiting distribution for
$(\bm{a}, \bm{\theta})$ may be non-trivial. Moreover, locating the global minimum of the least squares criterion usually requires a multi-dimensional grid
search over all possible values of the $s$ threshold parameters, which is typically computational infeasible. In fact, when $s$ is unknown,
\cite{gonzalo2002estimation} suggested a sequential estimation procedure for choosing $s$, under the homoscedasticity assumption and without the change plane
parameter $\bm{\theta}$. We are not aware of any results for more general models.

% In particular, for the $(s + 1)$-subgroup change-plane model (\ref{mcp}), when the number of thresholds $s$ and the change-plane parameter $\bm{\theta}$ in each regime are unknown, we proposed a iterative two-stage procedure to obtain the consistent number of change-planes, as well as to perform consistent parameter estimation.

We propose an iterative two-stage procedure for multi-threshold change plane estimation.  Given any consistent estimation $\hat{\bm{\theta}}$ in the first stage
we can obtain a consistent estimation of $s$ using a penalty-based change point detection algorithm. After we obtain $\hat{s}$, we can use the induced smoothing
approach introduced in section \ref{sec:scpl} to estimate $(\bm{\beta}, \bm{\delta}, \bm{\theta}, \bm{a})$ in the second stage. The details are as follows.

{\bf{The Splitting Stage.}} For a given estimator $\hat{\bm{\theta}}$, we denote $\hat W_i = \bm{Z}_i^{\trans}\hat{\bm{\theta}}$, $i = 1, \dots,n$. We then
generate the rank mapping $\{\iota_{(i)}: 1\leq i\leq n \}$ such that $\hat W_{\iota_{(i)}}$ is the $i$-th smallest value in $\{\hat W_i: 1\leq i\leq n\}$, and
can be arranged in the ascending order, that is,  $\hat W_{\iota_{(1)}}\leq \hat W_{\iota_{(2)}} \leq \dots \leq \hat W_{\iota_{(n)}}$. First we split the data
sequence into $q_n +1$ segments based on $\hat W_{\iota_{(i)}}$ where $q_n$ tends to infinity as $n\rightarrow\infty$. The data sequence is split such that the
first segment $\mathcal{I}_1 = \{i : \hat W_i \leq \hat W_{\iota_{(n - q_n m)} } \}$ involves $n - q_n m$ observations, and each of the other $q_n$ segments
$\mathcal{I}_j = \{i : \hat W_{\iota_{(n -(q_n -j+2)m)}} < \hat W_i \leq \hat W_{\iota_{(n -(q_n -j+1)m)}} \}$, $j = 2, . . . , q_n +1$ involves $m$
observations where $m = \lceil n /q_n \rceil$. %Define $b_j =  |\mathcal{I}_j|,~ j = 1, \dots, q_n+1$.

Let $\bm{Y}_{(j)} = (Y_i , i \in \mathcal{I}_j )^{\trans}$, $\bm{X}_{(j)} = (\bm{X}_i , i \in \mathcal{I}_j )^{\trans}$. Denote $\tilde{\bm{Y}} =
(\bm{Y}_{(1)}^{\trans}, \dots, \bm{Y}_{(q_n+1)}^{\trans} )^{\trans} $, $\tilde{\bm{X}} = (\bm{X}^{(1)} , \dots , \bm{X}^{(q_n+1)} )$ where $\bm{X}^{(1)} = (
\bm{X}_{(1)}^{\trans}, \dots , \bm{X}_{(q_n+1)}^{\trans})^{\trans} $ and $\bm{X}^{(j)} = (\bm{0}_{p\times \sum_{i=1}^{j-1}b_i}, $ $ {\bm{X}}_{(j)}^{\trans}, \dots,
{\bm{X}}^{\trans}_{(q_n + 1)})^{\trans}$, $j = 2, \dots, q_n+1$. The estimator $\tilde{\bm{\gamma}}^{*} = (\tilde{\bm{\beta}}^{\trans}_1,
\tilde{\bm{\delta}}^{\trans}_1, \dots, \tilde{\bm{\delta}}^{\trans}_{q_n})^{\trans}$ can be written as

\begin{eqnarray} \label{split}
\tilde{\bm{\gamma}}^{*}=\arg\min_{\bm{\gamma}} \left\{\frac{1}{n} \| \tilde{\bm{Y}} - \tilde{\bm{X}} \bm{\gamma}^{*} \|^2+ \sum_{j=1}^{q_n}
p_{\lambda_n}( \|\bm{\delta}_{j}\|)\right\},
\end{eqnarray}
We apply the group coordinate descent (GCD) algorithm to estimate $\tilde{\bm{\gamma}}^{*}$ from (\ref{split}). For simplicity, we write the estimator
$\tilde{\bm{\gamma}}^{*}=(\tilde{\bm{\gamma}}_1^{*\trans},\ldots,\tilde{\bm{\gamma}}^{*\trans}_{q_n+1})^{\trans}$. Denote $\hat{\mathcal{A}}=\{j:
\tilde{\bm{\gamma}}_j^{*}\neq 0, j=1,\ldots,q_n+1\}$, and
\begin{equation}\label{set.hats}
\hat{\mathcal{A}}^*=\{j:j\in \hat{\mathcal{A}},j-1\not\in \hat{\mathcal{A}},j=2,\ldots,q_n+1 \}= \{\hat k_1,\ldots,\hat k_{\hat s}\}, \quad\hat k_1<\ldots<\hat
k_{\hat s},
\end{equation}
which is a subset of $\hat{\mathcal{A}}$. It is obvious that if $j-1\not\in \hat{\mathcal{A}}$, $j\in \hat{\mathcal{A}}$ and $j+1\in \hat{\mathcal{A}}$, then
$j\in \hat{\mathcal{A}}^*$ and $j+1\not\in \hat{\mathcal{A}}^*$. Therefore, with each estimator $\hat{\bm{\theta}}$, we obtain the estimated number of change
planes $\hat{s} = |\hat{\mathcal{A}}^*|$. If the given estimator $\hat{\bm{\theta}}$ is consistent, then the estimated $\hat{s}$ in the splitting stage will
also convergence with high probability. The consistency of $\hat{s}$ can be guaranteed by Theorem \ref{thm1} in the next section. If $\hat{s} = 0$, we declare
there is no subgroup. If $\hat{s} > 0$, according to the proof of Theorem \ref{thm1}, the true threshold $a_{j}$ is highly likely to be located in $(\hat
W_{\iota_{(n-(q_n- \hat{k}_j +3)m)}},  \hat W_{\iota_{(n -(q_n - \hat{k}_j +1)m)}} ]$, $j = 1,\ldots , \hat{s}$. In the following step,  we can refine the
estimated thresholds $\hat{\bm{a}}$ and obtain all the regression coefficient estimates by an induced smoothing method.

{\bf{The Smoothed Refining Stage.}} Given an estimated number of change planes $\hat{s}$ from the previous stage, we can estimate the parameters  $\bm{a}$,
$\bm{\theta}$ and $\bm{\gamma} = (\bm{\beta}^{\trans}_1, \bm{\delta}^{\trans}_1, \dots, \bm{\delta}^{\trans}_{\hat{s}})^{\trans}$ in the model by minimizing the
following smoothed objective function
\begin{equation} \label{msmobj}
	 L_n^{*}(\bm{\eta}) = \frac{1}{n}\sum\limits_{i=1}^{n} \left\lbrack	Y_i - \bm{X}^{\trans}_i\bm{\beta}
-\bm{X}^{\trans}_i\sum_{k=1}^{\hat{s}}\bm{\delta}_{k}\Phi\left(\frac{\bm{Z}^{\trans}_i\bm{\theta} - a_k}{h}\right) \right\rbrack^2.	
\end{equation}
Write $\tilde{\bm{\eta}}^{*} =\arg\min_{\bm{\theta} \in \bm{\Theta}}\{L_n^{*}(\bm{\eta})\}$. For a non-sparse problem, to minimize (\ref{msmobj}) we can use the
familiar Newton-type algorithm. For a spares problem, similar to the single threshold change plane model, a penalty function can be added in (\ref{msmobj}) to
deal with the sparse parameterization among the coefficients $\bm{\gamma}$. Then $\bm{\eta}$ can be estimated by minimizing the following penalized objective
function
\begin{equation} \label{msmobjpen}
	Q_n^{*}(\bm{\eta}) = \frac{1}{n}\sum\limits_{i=1}^{n}\left\lbrack	Y_i - \bm{X}^{\trans}_i\bm{\beta}
-\bm{X}^{\trans}_i\sum_{k=1}^{\hat{s}}\bm{\delta}_{k}\Phi\left(\frac{\bm{Z}^{\trans}_i\bm{\theta} - a_k}{h}\right) \right\rbrack^2 +
p_{\lambda_n}(|\bm{\gamma}|).	
\end{equation}
where $p_{\lambda_n}(\cdot)$ is the penalty function. We consider SCAD and MC+ in the following numerical studies. Denote $\hat{\bm{\eta}}^{*}
=\arg\min_{\bm{\theta} \in \bm{\Theta}} \{Q_n^{*}(\bm{\eta})\} $, which can be obtained by an iterative penalized induced smoothing procedure similar to that
used in section 2.

We may repeat the splitting and smoothing stages many rounds until some convergence criterion is met. In particular, we terminate the iteration when the
estimated number of change planes remains unchanged. The detailed algorithm is described in the following:

\begin{itemize}
	\item {\sl Step 0}: Given an initial estimate of $\bm{\theta}$, say $\hat{\bm{\theta}}^{*}_{int}$, and set $\hat{\bm{\theta}}^{*} =
\hat{\bm{\theta}}^{*}_{int}/\|\hat{\bm{\theta}}^{*}_{int}\|$.
	\item {\sl Step 1}: Implement the Splitting stage. Minimizing (\ref{split}) and compute the index sets $\hat{\mathcal{A}}^*$ defined in (\ref{set.hats}),
and obtain the number of thresholds by $\hat{s} = |\hat{\mathcal{A}}^*|$.
	\item {\sl Step 2}: Given $\hat{s}$, update $(\hat{\bm{\theta}}^{*}, \hat{\bm{a}}^{*}, \hat{\bm{\gamma}}^{*})$ by minimizing objective function
(\ref{msmobj}) or (\ref{msmobjpen}) in the smoothed refining stage.
	\item {\sl Step 3}: Iterate Steps 1 and 2 until convergence.
\end{itemize}

\begin{remark}
The performance of splitting stage is dependent on the segment length $m$, and the selection of an optimal $m$ may follow the recommendation in
\cite{jin2015multi}. In the smoothed refining stage, the algorithm proposed in section 2 can be similarly adopted. The number of parameters could be quite
large, especially when we have a large number of subgroups. The inclusion of the penalty functions may lead to a sparse solution. The oracle property of the
estimated $\hat{\bm{\gamma}}^{*}$ will be given in Theorem \ref{thm3}. The tuning parameter $\lambda_n$ can be chosen by the BIC criterion under moderate- or
high-dimensional situations (\citealt{fan2013tuning}).
\end{remark}

\subsection{Asymptotic Properties}

In this section, we study the theoretical properties of the proposed estimation. To establish the asymptotic theory, we impose the following necessary
conditions to facilitate the technical proofs.

\begin{condition}\upshape \label{con.1}
 (a) $E(\bm{X}_i \bm{X}_i^{\trans}) = \Sigma_0$ is finite and positive definite.  $E(\bm{Z}_i\bm{Z}_i^{\trans})$ is positive definite. $\bm{Z}_i$ and
 $\epsilon_i$ are independent, $i=1, \dots,n$. $E(\epsilon_i| \bm{X}_i) = 0$ almost surely. (b) Let  $0<E\|(\bm{X}_i^{\trans}, \bm{Z}_i^{\trans})^{\trans}(\bm{X}_i^{\trans}, \bm{Z}_i^{\trans})\|^{\xi} < \infty$  , and $E\|(\bm{X}_i^{\trans}, \bm{Z}_i^{\trans})^{\trans}\epsilon_i\|^{\xi} < \infty$ for some $\xi > 1$.
 Furthermore, $E(\bm{X}_i \bm{X}_i^{\trans}|\bm{Z}_i) > 0$ almost surely.
\end{condition}

\begin{condition}\upshape\label{con.2}
	The parameter space for $\bm{\eta}$ is compact with $\min_{1\leq l< k \leq s}\{|a_{l}-a_{k}|\}$ and $\min \{\|\bm{\beta}/\sqrt{p} \|,$ $
\|\bm{\delta}_{1}/\sqrt{p}\|, \dots, \|\bm{\delta}_{s}/\sqrt{p} \|\}$ bounded away from zero.
\end{condition}

Let $\rho(t) = \lambda_n^{-1}p_{\lambda_n}(t)$ and $\bar{\rho}(t) = \rho^{'}(|t|)\sgn(t)$. We assume that the penalty function $p_{\lambda_n}(\cdot)$ satisfies
the following condition:
\begin{condition}\upshape\label{con.3}
	$p_{\lambda_n}(\cdot)$ is a symmetric function and it is nondecreasing and concave on $[0, \infty)$. There exists a constant $\nu > 0$ such that $\rho(t)$
is a constant for all $|t| \geq \nu \lambda_n$, and $\rho(0)=0$. $\rho^{'}(t)$ exists and is continuous except for a finite number of $t$, and $\rho^{'}(0+) =1$.

\end{condition}

Denote $\mathcal{I}_j^{0} = \{i : a_{j-1} < \bm{Z}^{\trans}_i\bm{\theta} \leq a_{j} \}$ for $j= 1, \ldots, s+1$, with true vector of threshold locations
$\bm{a}$ and change-plane $\bm{\theta}$. Similar to the definition of $\tilde{\bm{X}}$ and $\tilde{\bm{Y}}$, we define $\tilde{\bm{X}}_{\bm{a}}$ and
$\tilde{\bm{Y}}_{\bm{a}}$ by replacing $\mathcal{I}_j$ with $\mathcal{I}_j^{0}$. By Condition 1, we have
$\tilde{\bm{X}}_{\bm{a}}^{\trans}\tilde{\bm{X}}_{\bm{a}}/n \rightarrow_{a.s}\Upsilon $, where $\Upsilon$ is a positive definite matrix. To obtain the asymptotic
property of $\hat{\bm{\gamma}}^{*}$ in (\ref{msmobjpen}), we assume the following:
\begin{condition}\upshape\label{con.4}
	$\max_{u\geq 0}\{p_{\lambda_n}''(u)\}+\Lambda_{(s+1)p}(\Upsilon)>0$ where $\Lambda_{(s+1)p}(\Upsilon)$ is the smallest eigenvalue of $\Upsilon$.

\end{condition}

\begin{condition}\upshape\label{con.5}
	Let $W_{i} = \bm{Z}_i^{\trans}\bm{\theta}$, and $f_{W|V}(\cdot)$ denote the conditional density of $W_{i}$ given $\bm{V}_i = V$ and $f_W(\cdot)$ the density
of $W_{i}$, where $f_{W|V}(\cdot)$ is of compact support and has a bounded second derivative and $\bm{V}_i$ can be expressed as $(\bm{\delta}_{j}^{\trans}\bm{X}_i)^2$, $\bm{Z}_i \bm{Z}_i^{\trans} (\bm{\delta}_{j}^{\trans}\bm{X}_i)^2$ or $\bm{Z}_i(\bm{\delta}_{j}^{\trans}\bm{X}_i)^2$, respectively. $P(W_{i} \leq a_{j} ) = \tau_j$ where $0 < \tau_1 <
\ldots < \tau_s < 1$. Furthermore, $E(\epsilon^4_i|\bm{V}_i) < M$ for some $M<\infty$.
\end{condition}

\begin{condition} \upshape\label{con.6}
	$h\to 0$ and $nh^{2}\rightarrow 0$ as $n\to\infty$.
\end{condition}

\begin{remark}
   Condition \ref{con.1} for the design matrix is a common assumption (eg. Assumption $1$ of \cite{SEO2007704}) allowing for a regime specific
   heteroscedasticity. The error assumption can be relaxed to $\epsilon_i = \sigma(\bm{X}_i^{\trans} \bm{\beta})e_i$ where $e_i$ is independent with $\bm{X}_i$
   and $e_1 , \ldots , e_n$ are i.i.d. with mean zero and variance $\sigma^2$. Condition \ref{con.2} is about the parameter space, which excludes the
   possibility of a reduced model with less than $s+1$ subgroups by requiring $a_{j-1} < a_{j}$, $j=1, \ldots, s$. Conditions $3$ and Condition $4$ are often
   needed in shrinkage regression in high-dimensional data settings. The concave penalties such as MC+ and SCAD satisfy Condition \ref{con.3}. For the MC+
   penalty, Condition $4$ is equivalent to $\Lambda_{(s+1)p}(\Upsilon)>1/\nu$, and  for the SCAD penalty, Condition \ref{con.4} is equivalent to
   $\Lambda_{(s+1)p}(\Upsilon)>1/(\nu-1)$, which ensures the objective function (\ref{msmobjpen}) is globally convex. Condition \ref{con.5} is
   standard smoothing condition, see \cite{horowitz2002bootstrap} and \cite{SEO2007704}. Condition 5 also implies the existence of $s$ distinct jumps.
   Otherwise the model is non-identified.  Condition \ref{con.6} is to determine the rate for $h$.

\end{remark}

When $\bm{\theta}$ is either known or estimated consistently, we have $\hat W_i = W_{i} + o_p(1)$ where $W_{i} = \bm{Z}_i^{\trans}\bm{\theta}$. By law of large
numbers and Condition \ref{con.5}, we have $\sum_{i=1}^{n}\bm{1}(a_{j-1} <\hat W_i \leq a_{j})/n \rightarrow_{p} \tau_j - \tau_{j-1} >0$. Suppose that
$m\rightarrow \infty$, $m/n \rightarrow 0$. By Condition \ref{con.5}, it follows that with probability tending to $1$, $\sum_{i=1}^{n}\bm{1}(\hat W_{\iota_{(n
-(q_n -j+2)m)}} <\hat W_i \leq \hat W_{\iota_{(n -(q_n -j+1)m)}} )/n = m/n \rightarrow 0$. Thus there is at most one threshold located in each segment
$\mathcal{I}_j =\{\hat W_{\iota_{(n -(q_n -j+2)m)}} <\hat W_i \leq \hat W_{\iota_{ (n-(q_n -j+1)m)}} \}$ for large $n$ where $\hat W_{\iota_{ (n -(q_n -j+2)m)}
}$ and $\hat W_{\iota_{(n -(q_n -j+1)m)}}$, $j = 1, \ldots, q_n+1$ are defined in Section \ref{subsec:split}. Then a consistent estimation of the number of
change planes in the splitting stage can be guaranteed by the following theorem.

%Let $\hat \ma=\{j:\tilde{\bm{\gamma}}_j^* \neq 0, j=1,\ldots,q_n+1\}$, and  $\hat \ma^*$ be a subset of $\hat \ma$ such that $\hat \ma^*=\{j:j\in \hat\ma,j-1\not \in \hat \ma,j=2,\ldots,q_n+1\}=\{\hat k_{1},\ldots, \hat k_{\hat s}\}$, where $\hat{s}$ is the size of $\hat \ma^*$.

\begin{theorem}\label{thm1}
	 Suppose $m\rightarrow\infty$ and $m = O(n^r)$, where $0 < r \leq 1/2$ is a constant, $\lambda_n\rightarrow 0$ and $\lambda_n
\sqrt{n}/\log{n}\rightarrow\infty$ as $n\rightarrow\infty$. If Conditions 1-5 hold, then we have $\lim_{n\rightarrow\infty} P(\hat{s} = s) = 1$.
\end{theorem}

Let $\bm{\gamma} = (\gamma_{1},\ldots,\gamma_{(s+1)p})^{\trans}= (\bm{\beta}^{\trans},\bm{\delta}_{1}^{\trans},\ldots,\bm{\delta}_{s}^{\trans})^{\trans}$ be the
regression parameters in (\ref{mcp}) and $\mathcal{G}=\{j:\gamma_{j} \neq 0,j=1,\ldots,(s+1)p\}$ be the set of important variables in the model. For a given
consistent estimate $\hat{s}$, the consistency of smoothed least square estimator $\tilde{\bm{\eta}}^{*}$ which minimizing the unregularized objective function
(\ref{msmobj}) can be obtained by extending Theorem $1$ in \cite{SEO2007704} where $s=1$. We consider the estimator $\hat{\bm{\eta}}^{*}$ which minimizes the
penalized smooth objective function (\ref{msmobjpen}). The following theorem guarantees the consistency of our estimators. The proof is more complicated and
requires a detailed development.

\begin{theorem}\label{thm2}
	 Under Conditions 1-6, $\hat s=s$ and $\lambda_n\rightarrow 0$  as $n\rightarrow\infty$, there is a local minimizer $\hat{\bm{\eta}}^{*}$ of   $L_n^{*}(\bm{\eta})$
such that $\|\hat{\bm{\gamma}}^{*} -  \bm{\gamma}\| = O_p(\sqrt{1/n})$, $\|\hat{\bm{a}}^{*} -  \bm{a}\| = O_p(\sqrt{h/n})$ and $\|\hat{\bm{\theta}}^{*} -
\bm{\theta}\| = O_p(\sqrt{h/n})$, where $\|\hat{\bm{\theta}}^{*} \| = \|\bm{\theta}\|=1$.
\end{theorem}

We rewrite $\mathcal{G} = \{g_{1}, \ldots, g_{s+1} \}$, where $g_{j+1} = \{j_1, \ldots, j_{p_j} \}$ is the index set of $p_j$ nonzero covariates set in the
$j$th subgroup, $j = 0, 1, \ldots, s$. Without loss of generality, we shall write $\bm{\gamma}_p = (\bm{\gamma}_{(1)}^{\trans},
\bm{\gamma}_{(2)}^{\trans})^{\trans}$ to be a permuted version of $\bm{\gamma}$ where $\bm{\gamma}_{(1)} = (\bm{\gamma}_{g_1}^{\trans},\ldots,
\bm{\gamma}_{g_{s+1}}^{\trans} )^{\trans}$ with $\bm{\gamma}_{g_{j+1}} = (\gamma_{j_1},\ldots,\gamma_{j_{p_j}})^{\trans}$ and $\bm{\gamma}_{(2)}= \bm{0}$. For
$j=0, 1,\ldots ,s$, denote $\bm{X}_{i,g_{j+1}} = (X_{i,j_1},\ldots, X_{i,j_{p_j}})^{\trans}$.
 Denote  $\Sigma_1 = (\sigma_{1,jk})_{0\leq j,k \leq s}$ as the $\sum_{j=0}^{s}p_j \times \sum_{j=0}^{s}p_j$ block matrix, where the block $\sigma_{1,jk} = 4\sigma^2 E\bm{X}_{i,g_{j+1}}\bm{X}_{i,g_{k+1}}^{\trans}$ $\bm{1}(\bm{Z}_i^{\trans}\bm{\theta} > a_{j}\vee a_k)$, $\Sigma_2 = \frac{4}{nh}\mathop{\mathrm{diag}}\{\frac{\sigma^2}{2\sqrt{\pi}} A_j+\Pi \cdot B_j,j=2,\ldots,s\}$ be the $s\times s$ diagonal matrix where $A_j= E\{(\bm{\delta}_{j}^{\trans}\bm{X}_i)^2|\bm{Z}_i^{\trans}\bm{\theta}= a_{j} \}f_W(a_{j})$,
$B_j=E\{(\bm{\delta}_{j}^{\trans}\bm{X}_i)^4|\bm{Z}_i^{\trans}\bm{\theta} = a_{j} \}f_W(a_{j})$ and $\Pi = \int_{-\infty}^{\infty} \phi\left(s\right)^2\left(\bm{1}(s>0) -\Phi\left(s\right) \right)^2 d s$, and $\Sigma_3 = \frac{4}{nh}\sum_{j=1}^{s}  \left\{\frac{\sigma^2}{2\sqrt{\pi}} G_j +\Pi \cdot H_j \right\}$ be the $d\times d$ matrix where $G_j=E\left\{\bm{Z}_i\bm{Z}_i^{\trans}  (\bm{\delta}_{j}^{\trans} \bm{X}_i)^2|\bm{Z}_i^{\trans}\bm{\theta}=a_j \right\} f_{W}(a_j)$ and $H_j=E\left\{\bm{Z}_i\bm{Z}_i^{\trans}  (\bm{\delta}_{j}^{\trans} \bm{X}_i)^4|\bm{Z}_i^{\trans}\bm{\theta}=a_j \right\} f_{W}(a_j)$.

Let  $V_{11} = (v_{1k,l})_{0\leq k,l \leq s}$, where $v_{1k,l} = 2E\bm{X}_{i,g_{k+1}} \bm{X}_{i,g_{l+1}}^{\trans} \bm{1}(\bm{Z}_i^{\trans}\bm{\theta} > a_{k}\vee a_{l})$, $V_{22} = \mathop{\mathrm{diag}}(\frac{A_k}{\sqrt{\pi}}, k = 1, \dots,s)$, $V_{23} = (v_{23vk})_{1\leq v\leq d, 1\leq k\leq s}$
and $v_{23vk} = - \frac{1}{\sqrt{\pi}} E[Z_{iv}(\bm{\delta}_{g_{k+1}}^{\trans}\bm{X}_{i,g_{k+1}})^2|\bm{Z}_i^{\trans}\bm{\theta}=a_k]f_{W}(a_{k})$, and  $V_{33} = \frac{1}{\sqrt{\pi}}\sum_{j=1}^{s}\bm{G}_j$.

Denote
\begin{align*}
	\Gamma_{\lambda_n}  = &\mathrm{diag} \{p^{''}_{\lambda_n}(|\gamma_{0_1}|),\ldots, p^{''}_{\lambda_n}(|\gamma_{0_{p_0}}|), \ldots,
p^{''}_{\lambda_n}(|\gamma_{s_{1}}|), \ldots, p^{''}_{\lambda_n}(|\gamma_{s_{p_s}}|) \}.
\end{align*}
The limiting distributions of the estimators are developed in the following theorem.

\begin{theorem}\label{thm3}
	  Under Conditions 1-6, $\lambda_n \rightarrow0$ and $\lambda_n \sqrt{n}/\log{n}\rightarrow\infty$ as $n\rightarrow\infty$, with probability tending to $1$,
the penalized smooth estimator $\hat{\bm{\eta}}^{*} = (\hat{\bm{\gamma}}^{*}_{(1)}, \hat{\bm{\gamma}}^{*}_{(2)}, \hat{\bm{a}}^{*}, \hat{\bm{\theta}}^{*})$ in
Theorem \ref{thm2} satisfies
\begin{itemize}
	\item[(a)] Sparsity: $\hat{\bm{\gamma}}^{*}_{(2)} = \bm{0}$.
	\item[(b)] Asymptotic normality:
	\begin{align*}
		&\sqrt{n}(\hat{\bm{\gamma}}^{*}_{(1)} -  \bm{\gamma}_{(1)} ) \overset{D}{\to} N(0, (V_{11}+\Gamma_{\lambda_n})^{-1}\Sigma_1(V_{11}+\Gamma_{\lambda_n}
)^{-1}), \\
		%\[
        &\sqrt{n/h} \tilde{\bm{V}} \left(\begin{array}{c}
        \hat{\bm{a}}^{*} -  \bm{a} \\
        \hat{\bm{\theta}}^{*} -  \bm{\theta}
        \end{array}\right)\overset{D}{\to} N(0, \bm{\Omega} ).
        %\]
	\end{align*}
\end{itemize}
where $\tilde{\bm{V}}=
\begin{pmatrix}
V_{22} & V_{23} \\
\bm{P}_{\bm{\theta}}V_{23}^{\trans} & \bm{P}_{\bm{\theta}}V_{33}
\end{pmatrix}$, $\bm{\Omega} = \mathop{\mathrm{diag}}(\Sigma_2, \bm{P}_{\bm{\theta}}\Sigma_3\bm{P}_{\bm{\theta}} ) $, and $\bm{P}_{\bm{\theta}}= I-
\bm{\theta}\bm{\theta}^{\trans}$.
Furthermore, $\sqrt{n/h} \tilde{\bm{V}}\left(\begin{array}{l}
    \hat{\bm{a}}^{*} -  \bm{a} \\
    \hat{\bm{\theta}}^{*} -  \bm{\theta}
  \end{array}\right)$
and $\sqrt{n}(\hat{\bm{\gamma}}^{*}_{(1)} -  \bm{\gamma}_{(1)})$ are asymptotically independent.
\end{theorem}

 Theorem \ref{thm3} ensures that the penalized estimators enjoy the oracle property and work as well as when estimating $\hat{\bm{\gamma}}^{*}_{(1)},
 \hat{\bm{a}}^{*}, \hat{\bm{\theta}}^{*}$ with known $\hat{\bm{\gamma}}^{*}_{(2)} = \bm{0}$. Hence, our proposed MCPL estimation can be used to estimate
 parameters and select variables simultaneously without losing any efficiency.

%{ \color{red}
Theorem 3 may provide inference tools for many models simpler than ours but not studied in the literature yet. For example, it is interesting to consider the
case with one-dimensional thresholding variable $(i.e., d=1)$, where $\bm{\theta} = 1$ and $\bm{Z}_i^{\trans}\bm{\theta} = Z_i$. Then we can estimate
$\hat{\bm{\eta}}^{*}$ by the estimation method in this paper, and obtain the distribution theory of the resulting estimator in the following corollary.

\begin{corollary}\label{corol.onestep}
	 Suppose Conditions $1$-$6$ hold, we have $\lim_{n\rightarrow\infty} P(\hat{s} = s) = 1$ and furthermore $\sqrt{n}(\hat{\bm{\gamma}}^{*} -  \bm{\gamma})$
and $\sqrt{n/h}(\hat{\bm{a}}^{*} -  \bm{a})$ are asymptotically independent, and $$\sqrt{n}(\hat{\bm{\gamma}}^{*}_{(1)} -  \bm{\gamma}_{(1)}) \overset{D}{\to}
N(0, (V_{11}- \Gamma_{\lambda_n})^{-1}\Sigma_1(V_{11}-\Gamma_{\lambda_n} )^{-1}),$$ $$\sqrt{n/h}(\hat{\bm{a}}^{*} -  \bm{a}) \overset{D}{\to} N(0,
V_{22}^{-1}\Sigma_2 V_{22}^{-1}).$$
\end{corollary}
We note that Li and Jin (2018) provided consistency results for such estimators but did not present the asymptotic distribution theory. This corollary offers a
complement to their results.

The proofs of all the theorems are given in the supplementary materials of this paper.

%}

\section{Simulation Studies}
\label{sec:sim}

We conducted extensive simulation studies to investigate the empirical performance of the proposed method for subgroup detection and the estimation for the
change-plane parameters. We consider the following examples to compare the performance of our methods. Specifically, For all cases the random noise $\epsilon$
is normally distributed with mean zero and variance $0.25$. We generate the regressors $\bm{X}_i = (X_{i1}, \ldots,  X_{ip})^{\trans}$ with an intercept
$X_{i1}=1$ and $(X_{i2}, \ldots,  X_{ip})^{\trans} \sim N(0, \Sigma)$, for different structures of covariance matrix $\Sigma=(\Sigma_{ij})$:
\begin{itemize}
	\item [(1)] $\Sigma_1$: $\Sigma_{ij} = \bm{1}_{\{i=j\}}$ for all $i$, $j$ (the identity matrix);
	\item [(2)] $\Sigma_2$: $\Sigma_{ij} = 0.5^{|i-j|}$ for all $i$, $j$ (Toeplitz matrix);
	\item [(3)] $\Sigma_3$: $\Sigma_{ij} = 1-0.5\cdot\bm{1}_{\{i\neq j\}}$ for all $i$, $j$ (equi-correlation).
\end{itemize}
We choose the threshold variables $\bm{Z}$ to be a subset of $\bm{X}$. Specifically, we consider the following examples:

\textbf{Example 1}: (Single threshold) We consider the single threshold change plane model (\ref{scp}) with $p = 6$ and $d = 2$, and we choose sample size
$n=150$ and $n=300$. We specify the true baseline coefficients $\bm{\beta} = (2,1,1,1,1,1)^{\trans}$, the enhanced effects in the subgroup $\bm{\delta} = (-1,
0, 0, -1, -1, -1)^{\trans}$, then $\bm{\gamma} = (\gamma_{1}, \ldots, \gamma_{12} )^{\trans} = (\bm{\beta}^{\trans}, \bm{\delta}^{\trans})^{\trans}$. Let the
threshold variables be $\bm{Z}_i = (1, X_{i1}, X_{i2})^{\trans}$ with the first element be the constant $1$, the true change-plane parameter is chosen as
$\bm{\theta} = (-0.15, 0.3, 0.942)^{\trans}$.

\textbf{Example 2}: (Multi-threshold) We consider a multiple threshold change plane model (\ref{mcp}) with two thresholds ($s = 2$). We choose sample size
$n=150, 300, 500$, and $p=5, 20$, and specify the true baseline coefficients $\bm{\beta} = (2,1,1,1,1,1, \underbrace{0, \dots, 0}_{p-5})^{\trans}$, the enhanced
treatment effect in the subgroup $\bm{\delta} = (\bm{\delta}_1^{\trans}, \bm{\delta}_2^{\trans})^{\trans}$ where $\bm{\delta}_1 = (-1,0,0,-1, -
1,$ $\underbrace{0, \dots, 0}_{p-5})^{\trans}$ and $\bm{\delta}_2 = (0,-1, 1, 0, 0, 0, \underbrace{0, \dots, 0}_{p-5})^{\trans}$, then $\bm{\gamma} =
(\gamma_{1}, \ldots, \gamma_{3\times p} )^{\trans} = (\bm{\beta}^{\trans}, \bm{\delta}^{\trans})^{\trans}$. Choose the threshold variables as $\bm{Z}_i =
(X_{i2}, X_{i3}, X_{i4})^{\trans}$ and the true change-plane parameter is chosen as $\bm{\theta} = (0.75, -0.25, 0.612)^{\trans}$, where true thresholds $a_1 =
-0.524$, $a_2 = 0.253$, which correspond to the 30\% and 60\% lower percentiles of the standard normal distribution. This scenario generates roughly the same
number of subjects in the three subgroups.

\textbf{Example 3}: (No subgroup) The same as Example $2$ except $n=300$, $s = 0$ and $\bm{\beta} = (1, 0, 2, 0, 0, 0)$.

\textbf{Example 4}: (Unequal group sizes) The same as Example $2$ except true thresholds $a_1 = -\sqrt{2}/2$, $a_2 = \sqrt{2}/2$ which generates unequal
sample size in the subgroups.

All results for the examples are based on 500 simulations and reported in Tables 1 to 10. In all tables, ``Bias'' denotes the estimation bias, ``SD'' is the
empirical standard deviation of the estimates parameters, and ``RMSE'' is the root of the mean square errors. In addition, to measure how close the estimated
grouping structure approaches the true one, we introduce the normalized mutual information (NMI), which is a common measure for similarity between clusterings
(\citealt{ana2003robust}). Suppose $\mathbb{C} = {C_1 , C_2 ,\ldots }$ and $\mathbb{D} = {D_1 , D_2 , \ldots }$ are two sets of disjoint clusters of $\{1,\ldots
,n\}$, define
$$ \textup{NMI}(\mathbb{C}, \mathbb{D}) = \frac{I(\mathbb{C}, \mathbb{D})}{[H(\mathbb{C})+ H(\mathbb{D})]/2}  $$
where $I(\mathbb{C}, \mathbb{D}) = \sum_{k,j} (|C_k \cap D_j |/n) \log(n |C_k \cap D_j|/|C_k ||D_j|)$ is the mutual information between $\mathbb{C}$ and
$\mathbb{D}$, and $H(\mathbb{C}) = -\sum_k (|C_k|/n) \log(|C_k |/n)$ is the entropy of $\mathbb{C}$. NMI$(\mathbb{C}, \mathbb{D})$ takes values on $[0, 1]$, and
larger NMI implies the two groupings are closer. In particular, NMI $= 1$ means that the two groupings are exactly the same.

Table \ref{exp1} and \ref{exp1.theta.hat} present the bias, SD and root of the mean square errors (RMSE) for the estimated coefficients and the change-plane
parameters using our proposed methods under Example $1$. We can see that the estimated parameters are all very close to the true values. To assess the
performance of variable selection, Table \ref{exp1.no0} shows the number of correctly selected zeros and incorrectly selected zeros in
$\tilde{\bm{\gamma}}^{*}$. We can see that our estimators can identify the true sparse structure accurately.

\begin{table}[h]\scriptsize
\begin{center}
\caption{Simulation results for Example 1. Bias is the average of estimated parameter minus the true value. SD is the empirical standard deviation. RMSE refers
to the relative mean squared errors.}
\label{exp1}
\centerline{
\begin{tabular}{lllrrrrrrrrrr}
\hline
  &&& $\beta_0$ &$\beta_1$& $\beta_2$& $\beta_3$ & $\beta_4$&$\beta_5$&$\delta_0$&$\delta_3$&$\delta_4$&$\delta_5$ \\
  \hline
 $\Sigma_{ij} = \bm{1}_{\{i=j\}}$ & $n=150$ & Bias & -0.006 & -0.006 & -0.008 &-0.003 &-0.003&      -0.001&0.023& 0.006& 0.009  &0.004 \\
  & &  SD & 0.091 &0.049 &0.080 &0.064 &0.059 &      0.065&0.178 & 0.109& 0.102 &0.100 \\
  & &  RMSE & 0.091 &0.050& 0.081 &0.064 &      0.059& 0.065 &0.179& 0.109 &0.102& 0.101\\
  \cline{2-13}
  & $n=300$ &Bias& -0.009&      -0.002 & -0.009 & -0.001& -0.003& -0.003  &0.020& 0.003& 0.006 &0.010   \\
  && SD&0.056 &0.033 &0.049& 0.038 & 0.038 &0.038 &0.101& 0.059 &0.061& 0.057 \\
  && RMSE& 0.057 &0.033 &0.050 &0.038 &0.038 &0.038&0.103& 0.059 &0.061 &0.058 \\
  \hline
  $\Sigma_{ij} = 0.5^{|i-j|}$ & $n=150$ & Bias & -0.006 &-0.004 &-0.007& 0.004& -0.009& 0.003& 0.017& -0.001& 0.016& -0.001 \\
  & & SD & 0.094 &0.054& 0.085 &0.077&0.078&  0.072&0.181& 0.113& 0.119& 0.106 \\
  & & RMSE& 0.094& 0.054 &0.086&0.077&0.078&   0.072&0.182&0.113& 0.120& 0.106 \\
  \cline{2-13}
  & $n=300$ & Bias& -0.004&0.001      &-0.007& 0.001 &0.002 &-0.008&0.013& 0.003&  -0.002&  0.014\\
  && SD & 0.056 &0.038 &0.054 &0.052&0.053& 0.047&0.105& 0.075& 0.076& 0.069 \\
  && RMSE & 0.056&0.038 &0.055&0.052&0.053&0.048& 0.105& 0.075& 0.076& 0.071 \\
  \hline
  $\Sigma_{ij} = 1-0.5\cdot\bm{1}_{\{i\neq j\}}$ & $n=150$ & Bias& -0.009 &-0.002& -0.004 &-0.001& -0.007 &-0.003&-0.011& 0.011& -0.007&  0.003  \\
  & & SD & 0.094&0.062&0.085 &0.080&0.075 &0.082&0.188& 0.117& 0.112& 0.119   \\
  && RMSE& 0.094 &0.062&0.085 &0.080 &0.075& 0.082&0.189& 0.118& 0.112& 0.119 \\
  \cline{2-13}
  & $n=300$ & Bias & -0.003&      -0.002 &-0.003& 0.002& -0.005 &-0.002&0.007& -0.001&  0.004&  0.005\\
  && SD & 0.059&0.045&0.051&0.055 &0.052& 0.054&0.105& 0.078& 0.075& 0.081  \\
  && RMSE&0.059 &0.045&0.051 &0.055 &0.052& 0.054&0.106& 0.078& 0.076& 0.082  \\
  \hline
\end{tabular}}
\end{center}
\end{table}

\begin{table}[h]\footnotesize
\begin{center}
\caption{Estimation performance for the change-plane estimation for Example $1$. Bias is the average of estimated parameter minus the true value. SD is the
empirical standard deviation. RMSE refers to the relative mean squared errors.}
\label{exp1.theta.hat}
\begin{tabular}{llccccccccc}
\hline
  && \multicolumn{3}{c}{$\theta_1$} &\multicolumn{3}{c}{$\theta_2$}& \multicolumn{3}{c}{$\theta_3$} \\
  \hline
  & &  Bias &SD & RMSE & Bias & SD& RMSE & Bias & SD& RMSE\\
  \hline
 $\Sigma_{ij} = \bm{1}_{\{i=j\}}$ & $n=150$ &0.003& 0.038& 0.038& -0.007& 0.049& 0.050& 0.005& 0.022& 0.023 \\
  \cline{2-11}
  & $n=300$ & -0.001& 0.015& 0.015& -0.001& 0.022& 0.022& 0.001& 0.007& 0.007\\
  \hline
  $\Sigma_{ij} = 0.5^{|i-j|}$ & $n=150$ &-0.001& 0.035& 0.035& -0.010& 0.057& 0.058& 0.006& 0.025& 0.025 \\
  \cline{2-11}
  & $n=300$ &0.001& 0.016& 0.016& -0.001& 0.028& 0.028& 0.001& 0.012& 0.012 \\
  \hline
 $\Sigma_{ij} = 1-0.5\cdot\bm{1}_{\{i\neq j\}}$ &$n=150$ &  0.004& 0.043& 0.043& -0.013& 0.067& 0.068& 0.009& 0.033& 0.034  \\
  \cline{2-11}
  & $n=300$& -0.002& 0.024& 0.024& -0.004& 0.039& 0.039& 0.003& 0.018& 0.019\\
  \hline
\end{tabular}
\end{center}
\end{table}

\begin{table}[h]
\begin{center}
\caption{Variable selection results for Example $1$. In this case two coefficients are zero. Correct and Incorrect refer to the average number of estimated zero
coefficients corresponding to zero and non-zero coefficients, respectively.}
\label{exp1.no0}
\begin{tabular}{llcc}
\hline
  && \multicolumn{2}{c}{Avg. no. of $0$ coefficients}\\
  \hline
  & &  Correct & Incorrect \\
  \hline
 $\Sigma_{ij} = \bm{1}_{\{i=j\}}$ & $n=150$ & 1.996& 0.018 \\
  & $n=300$ & 1.998 & 0\\
  \hline
  $\Sigma_{ij} = 0.5^{|i-j|}$ & $n=150$ & 1.996 & 0.008\\
  & $n=300$ & 1.996 &0 \\
  \hline
  $\Sigma_{ij} = 1-0.5\cdot\bm{1}_{\{i\neq j\}}$ &$n=150$ & 1.990& 0.016 \\
  & $n=300$& 1.996 & 0 \\
  \hline
\end{tabular}
\end{center}
\end{table}

For multiple threshold change plane models under Example 2 and 4 and no subgroup model under Example 3, the estimation results for the number of thresholds
$\hat{s}$ are reported in Table \ref{exp23} and \ref{exp45} based on $500$ simulations, respectively. Our methods can correctly identify the number of
thresholds with very high probability in both cases.

\begin{table}[h]
\begin{center}
\caption{Frequency of estimated $\hat{s}$ in 500 simulations for Examples 2 and 3.}
\label{exp23}
\begin{tabular}{llllccccc}
  \hline
  &\multicolumn{3}{c}{$\hat{s}$} &0& 1& 2 &3&4 \\
  \hline
 $\Sigma_{ij} = \bm{1}_{\{i=j\}}$ &$s = 2$ & $n=150$ &$p=5$&0&0&488&11&1\\
  & & & $p=20$  &38 &188 &271&3&0  \\
  %\cline{3-9}
  && $n=300$ &$p=5$&0 &0 & 491&9&0 \\
  && & $p=20$ &0&21& 476&3 & 0 \\
  %\cline{3-9}
  && $n=500$ &$p=5$ &0 &0&487&13&0 \\
  && & $p=20$  &0&0&493&7&0  \\
    \cline{2-9}
  &$s = 0$ & $n=300$ &$p=5$ &500&0&0&0&0  \\
  && & $p=20$  &499&1&0&0&0 \\
  \hline
  $\Sigma_{ij} = 0.5^{|i-j|}$ & $s = 2$ & $n=150$ &$p=5$ &0&1&489&10&0 \\
  & & & $p=20$  & 6&294&198&2&0  \\
  %\cline{3-9}
  && $n=300$ &$p=5$ &0 &0 & 484&16&0\\
  && & $p=20$ &0&39& 454&7 & 0 \\
  %\cline{3-9}
  && $n=500$ &$p=5$ &0 &0&483&17&0 \\
  && & $p=20$  &0&0&490&10&0  \\
  \cline{2-9}
  &$s = 0$ & $n=300$ &$p=5$ &500&0&0&0&0  \\
  && & $p=20$  &499&1&0&0&0  \\
  \hline
  $\Sigma_{ij} = 1-0.5\cdot\bm{1}_{\{i\neq j\}}$ & $s = 2$ & $n=150$ &$p=5$ &0&8&452&39&1 \\
  & & & $p=20$  & 3&367 &127&3&0  \\
  %\cline{3-9}
  && $n=300$ &$p=5$ &0&0&446&52&2  \\
  && & $p=20$ & 0&117&346&37&0\\
  %\cline{3-9}
  && $n=500$ &$p=5$ &0 &0&432&67&1 \\
  && & $p=20$  &0&1&431&68&0  \\
  \cline{2-9}
  & $s = 0$ & $n=300$ &$p=5$ &500&0&0&0&0\\
  && & $p=20$  &500&0&0&0&0 \\
  \hline
\end{tabular}
\end{center}
\end{table}

\begin{table}[h]
\begin{center}
\caption{Frequency of estimated $\hat{s}$ in 500 simulations for Example 4.}
\label{exp45}
\begin{tabular}{lllccccccc}
  \hline
  &\multicolumn{2}{c}{$\hat{s}$} &0& 1& 2 & 3&4 &5&6\\
  \hline
$\Sigma_{ij} = \bm{1}_{\{i=j\}}$ &  $n=150$ &$p=5$ &0&1 &493&6&0&0&0\\
  & &  $p=20$  &86&89&306&19&0&0&0  \\
  %\cline{3-11}
  & $n=300$ &$p=5$&0 &0 &492&8&0&0&0\\
  & & $p=20$ &0& 2& 488&10&0&0&0  \\
  %\cline{3-11}
  & $n=500$ &$p=5$ &0 &0&492&6&2&0&0 \\
  & & $p=20$  &0&0&497&3&0&0&0  \\
  \hline
  $\Sigma_{ij} = 0.5^{|i-j|}$ &$n=150$ &$p=5$ & 0 & 2&490&8&0&0&0\\
  &  & $p=20$  & 11&257&231&1&0&0&0  \\
  %\cline{3-11}
  & $n=300$ &$p=5$ &0 &0 &494&5&0&0&1 \\
  & & $p=20$ &0& 13& 479&8&0&0&0\\
  %\cline{3-11}
  &$n=500$ &$p=5$ &0&0&489&11&0&0&0  \\
  & & $p=20$  &0&0&487&13&0&0&0  \\
  \hline
  $\Sigma_{ij} = 1-0.5\cdot\bm{1}_{\{i\neq j\}}$  & $n=150$ &$p=5$ &0&1&491&8&0&0&0 \\
   & & $p=20$  &12 &300 &187&1&0&0&0  \\
  %\cline{3-11}
  & $n=300$ &$p=5$ &0&0&471&29&0&0&0  \\
  & & $p=20$ & 0&34&458&8&0&0&0\\
  %\cline{3-11}
  & $n=500$ &$p=5$ &0&0&457&43&0&0&0 \\
  & & $p=20$  & 0 &0&458&42&0&0&0  \\
  \hline
\end{tabular}
\end{center}
\end{table}

\begin{figure}[htb!]
\begin{center}
{\tiny
\centerline{\includegraphics[scale=0.28]{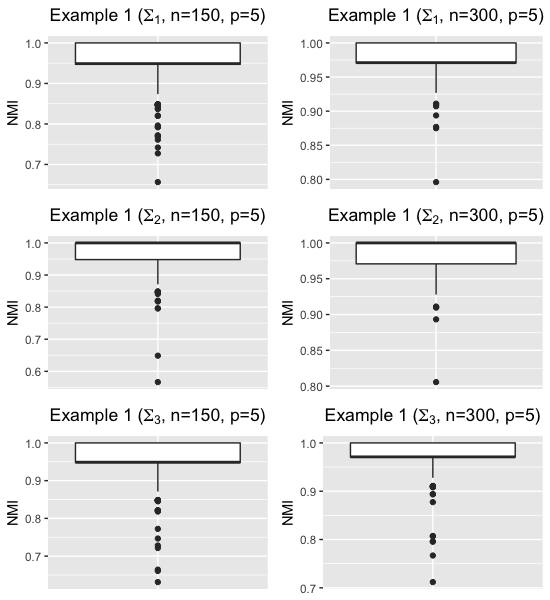}
\includegraphics[scale=0.28]{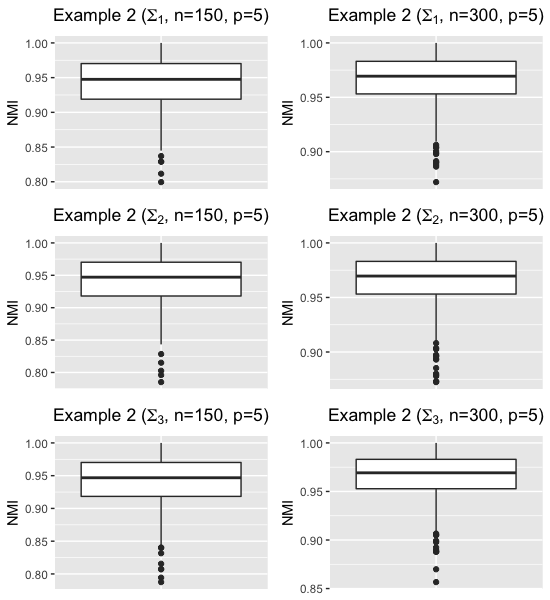}
\includegraphics[scale=0.28]{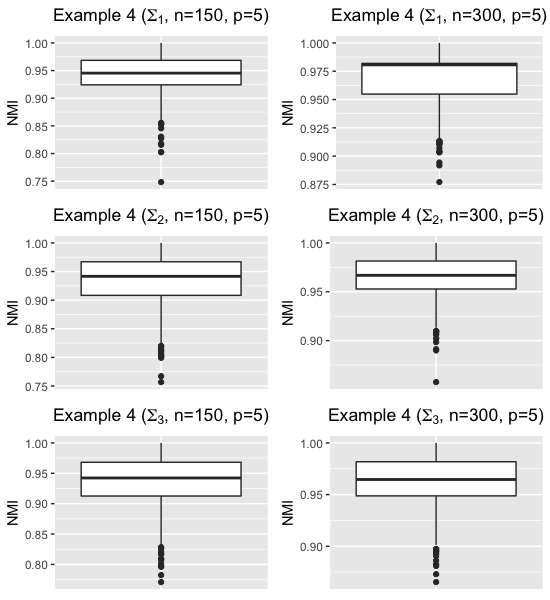}}}

\caption{The normalized mutual information (NMI) in Example 1, 2 and 4, where $n=150,300$, and $p=5$.}\label{fig:nmi}
\end{center}
\end{figure}

\begin{figure}[htb!]
\begin{center}
{\tiny
\centerline{
\includegraphics[scale=0.45]{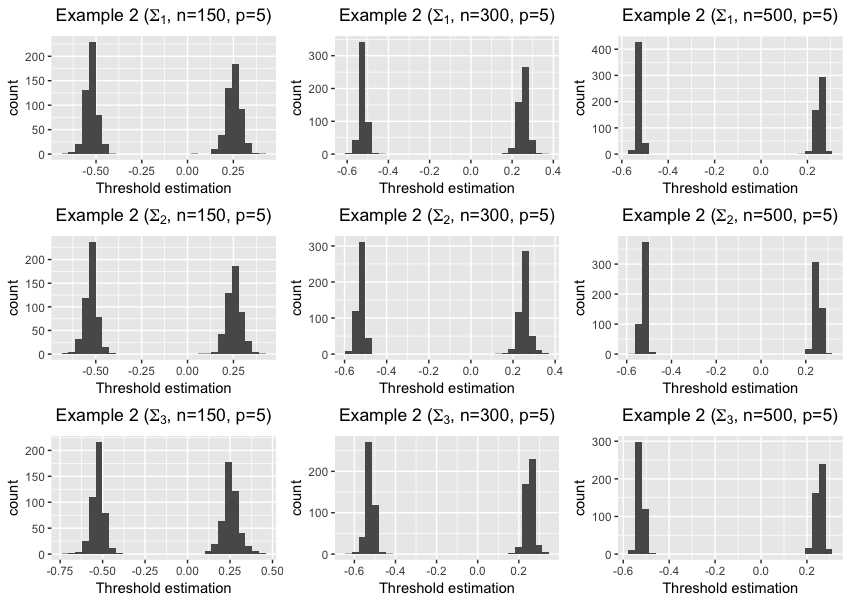}}}
\end{center}
\begin{center}
\vspace{-.4in}
{\tiny
\centerline{
\includegraphics[scale=0.45]{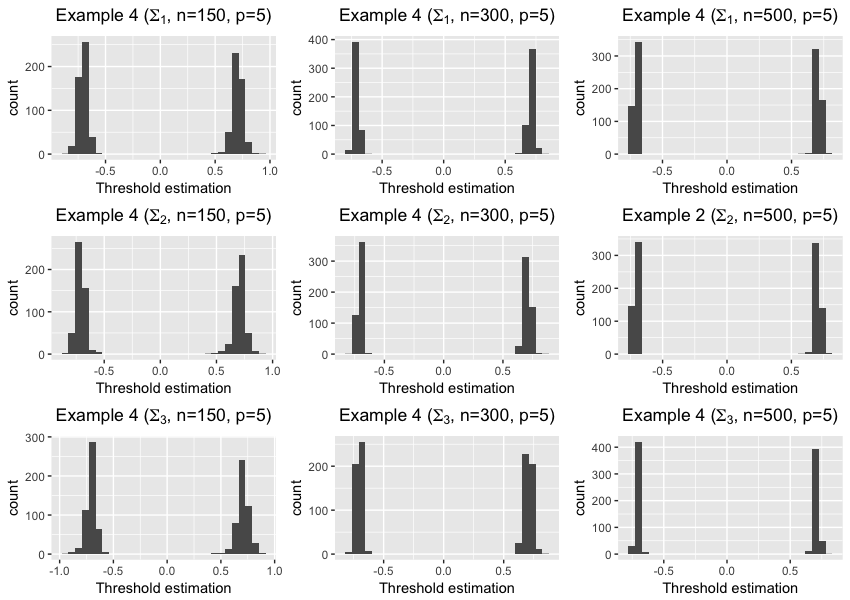}}}
\caption{Histograms of the estimated thresholds for Example $2$ and $4$. The true values are $(-0.524, 0.253)$ and $(-\sqrt{2}/2, \sqrt{2}/2)$ for Example $2$
and $4$ respectively.}\label{fig:exp24.a}
\end{center}
\end{figure}

Figure \ref{fig:nmi} shows boxplots of NMI for Example $1$, $2$ and $4$. We observe that our estimation enjoys a high agreement with the true group structure in
both single threshold and multiple threshold change plane models. Figure \ref{fig:exp24.a} displays the histograms of the estimated thresholds for Example $2$
and $4$ respectively, indicating the empirical estimates are very close to and symmetrically distributed around the true change points.

\begin{table}[h]
\begin{center}
\caption{Estimation performance for the threshold estimation for Example $2$. Bias is the average of estimated parameter minus the true value. SD is the
empirical standard deviation. RMSE refers to the relative mean squared errors.}
\label{exp2.a.hat}
\begin{tabular}{lllcccccc}
 \hline
 & & & \multicolumn{3}{c}{ $a_1$ } &  \multicolumn{3}{c}{$a_2$} \\
    \cline{4-9}
      & & & Bias &SD & RMSE & Bias & SD& RMSE \\
    \hline
  $\Sigma_{ij} = \bm{1}_{\{i=j\}}$ &$n=150$ & $p=5$ &0.001 &0.034 &0.033&0.001 &0.041&0.041  \\
   & & $p=20$ & 0.005 & 0.035 &0.035 &-0.003&0.047 &0.047\\
  & $n=300$& $p=5$ &-0.001 & 0.017& 0.017 & 0.001 & 0.022 & 0.022  \\
  && $p=20$&0.001&0.019&0.019&0.001&0.021&0.021\\
  & $n=500$ & $p=5$ & 0.001 &0.009 &0.010& 0.001&0.013&0.013\\
  && $p=20$ & 0.001&0.011&0.011&0.001&0.012&0.012\\
  \hline
  $\Sigma_{ij} = 0.5^{|i-j|}$ & $n=150$ &  $p=5$& 0.001& 0.033&0.033&-0.001&0.041&0.041 \\
  && $p=20$ &0.008&0.044&0.045&-0.006&0.048&0.048 \\
  & $n=300$ & $p=5$ & -0.001 & 0.017& 0.017 & 0.001 & 0.024 & 0.024  \\
  && $p=20$ & 0.001 & 0.019 &0.019 &-0.001 & 0.022 &0.022\\
  & $n=500$ & $p=5$ & -0.001 &0.010 &0.010& 0.001&0.014&0.014\\
  && $p=20$ & -0.001 & 0.010 &0.010 &-0.001& 0.012&0.012\\
  \hline
  $\Sigma_{ij} = 1-0.5\cdot\bm{1}_{\{i\neq j\}}$ &$n=150$ & $p=5$&0.003&0.037&0.038&-0.001&0.049&0.049 \\
  && $p=20$ & 0.012 &0.047&0.049&-0.007&0.049&0.050\\
  & $n=300$ & $p=5$ &-0.001 & 0.021& 0.021 & 0.002 & 0.022 & 0.022  \\
  && $p=20$ & 0.002 & 0.021&0.021&-0.001&0.025&0.025\\
  &$n=500$ & $p=5$ & -0.001 &0.012 &0.012& -0.001&0.015&0.015\\
  && $p=20$&-0.001&0.013&0.013&0.001&0.015&0.015\\
  \hline
\end{tabular}
\end{center}
\end{table}

\begin{table}[h]\footnotesize
\begin{center}
\caption{Estimation performance for the change-plane estimation for Example $2$. Bias is the average of estimated parameter minus the true value. SD is the
empirical standard deviation. RMSE refers to the relative mean squared errors.}
\label{exp2.theta.hat}
\centerline{
\begin{tabular}{lllccccccccc}
 \hline
& & & \multicolumn{3}{c}{ $\theta_1$ } &  \multicolumn{3}{c}{$\theta_2$} & \multicolumn{3}{c}{$\theta_3$}\\
    \cline{4-12}
     & & & Bias & SD& RMSE & Bias&SD & RMSE& Bias&SD & RMSE \\
    \hline
  $\Sigma_{ij} = \bm{1}_{\{i=j\}}$ & $n=150$ &  $p=5$ &0.001 &0.018 &0.017&0.001 &0.027&0.027&0.001 &0.020&0.020  \\
  & &  $p=20$ & 0.001 & 0.017& 0.017 &0.000& 0.027& 0.026& 0.001& 0.020& 0.020\\
  & $n=300$ &$p=5$ &-0.001 & 0.009& 0.009 & 0.001 & 0.013 & 0.013 & 0.001 & 0.011 &0.011 \\
  &&$p=20$ &-0.001& 0.011& 0.011& -0.001& 0.014& 0.014& 0.001& 0.012& 0.012 \\
  &$n=500$ &$p=5$ &0.001& 0.006& 0.006& 0.001& 0.008& 0.008& 0.001& 0.006& 0.006 \\
  &&$p=20$&0.001& 0.006& 0.006& 0.001& 0.008& 0.008& 0.001& 0.006& 0.006 \\
  \hline
  $\Sigma_{ij} = 0.5^{|i-j|}$ & $n=150$ &  $p=5$ &0.001 &0.020&0.020&-0.003&0.028&0.029&-0.001&0.022&0.022\\
  &&$p=20$& 0.004& 0.024& 0.025& 0.001& 0.032& 0.032& -0.003& 0.028& 0.029\\
  & $n=300$ &  $p=5$ & 0.001& 0.010& 0.010& -0.001& 0.014& 0.014& 0.001& 0.011& 0.011\\
  &&$p=20$ & 0.001& 0.011& 0.011& 0.001& 0.014& 0.014& 0.001& 0.012& 0.012  \\
  & $n=500$ &  $p=5$ &0.001& 0.007& 0.007& 0.001& 0.009& 0.009& 0.001& 0.007& 0.007 \\
  &&$p=20$&0.001& 0.007& 0.007& 0.001& 0.008& 0.008& 0.001& 0.007& 0.007  \\
  \hline
  $\Sigma_{ij} = 1-0.5\cdot\bm{1}_{\{i\neq j\}}$ & $n=150$ &  $p=5$ & 0.001& 0.028& 0.028& -0.002& 0.028& 0.028& 0.001& 0.032& 0.032 \\
  &&$p=20$ &0.003& 0.046& 0.046& -0.007& 0.036& 0.037& -0.002& 0.046& 0.046 \\
  & $n=300$ &  $p=5$ & 0.001& 0.016& 0.016& -0.001& 0.016& 0.016& 0.001& 0.018& 0.018\\
  &&$p=20$ &0.001& 0.016& 0.016& 0.001& 0.017& 0.017& 0.001& 0.018& 0.018 \\
  & $n=500$ &  $p=5$ &0.001& 0.010& 0.010& 0.001& 0.009& 0.009& 0.001& 0.011& 0.011 \\
  && $p=20$ &0.001& 0.009& 0.009& 0.001& 0.009& 0.009& 0.001& 0.010& 0.010 \\
  \hline
\end{tabular}}
\end{center}
\end{table}

\begin{table}[h]
\begin{center}
\caption{Estimation performance for the threshold estimation for Example $4$. Bias is the average of estimated parameter minus the true value. SD is the
empirical standard deviation. RMSE refers to the relative mean squared errors.}
\label{exp4.a.hat}
\begin{tabular}{lllcccccc}
 \hline
 & & & \multicolumn{3}{c}{ $a_1$ } &  \multicolumn{3}{c}{$a_2$} \\
    \cline{4-9}
      & & & Bias &SD & RMSE & Bias & SD& RMSE \\
    \hline
  $\Sigma_{ij} = \bm{1}_{\{i=j\}}$ & $n=150$ & $p=5$ & -0.001&0.040&0.040&0.002&0.049&0.049 \\
   & & $p=20$ &0.006&0.049&0.05&0.001&0.046&0.046\\
  & $n=300$ & $p=5$ &-0.002&0.019&0.019&0.001&0.024&0.024 \\
  && $p=20$&0.001&0.020&0.020&0.001&0.025&0.025  \\
  & $n=500$ & $p=5$ &0.001&0.012&0.012&0.001&0.017&0.017 \\
  && $p=20$ &0.001&0.011&0.011&0.001&0.014&0.014 \\
  \hline
  $\Sigma_{ij} = 0.5^{|i-j|}$ & $n=150$ &  $p=5$&0.003&0.041&0.042&0.004&0.055&0.055  \\
  && $p=20$ &0.014&0.050&0.052&0.006&0.048&0.049  \\
  & $n=300$ & $p=5$ &0.002&0.020&0.020&0.001&0.028&0.028  \\
  && $p=20$ & 0.001&0.022&0.022&-0.001&0.028&0.028 \\
  & $n=500$ & $p=5$ &0.001&0.012&0.012&0.001&0.017&0.017 \\
  && $p=20$ & 0.001&0.013&0.013&0.001&0.017&0.017\\
  \hline
  $\Sigma_{ij} = 1-0.5\cdot\bm{1}_{\{i\neq j\}}$ & $n=150$ & $p=5$& 0.001&0.047&0.047&0.002&0.058&0.058 \\
  && $p=20$ & 0.012&0.057&0.058&0.005&0.065&0.065\\
  & $n=300$ & $p=5$ & 0.001&0.024&0.024&0.001&0.030&0.030  \\
  && $p=20$ &0.001&0.026&0.026&-0.001&0.026&0.026 \\
  & $n=500$ & $p=5$ & 0.001&0.014&0.014&0.001&0.018&0.018\\
  && $p=20$&-0.001&0.014&0.014&0.001&0.018&0.018 \\
  \hline
\end{tabular}
\end{center}
\end{table}

\begin{table}[h]\footnotesize
\begin{center}
\caption{Estimation performance for the change-plane estimation for Example $4$. Bias is the average of estimated parameter minus the true value. SD is the
empirical standard deviation. RMSE refers to the relative mean squared errors.}
\label{exp4.theta.hat}
\centerline{
\begin{tabular}{lllccccccccc}
 \hline
 & & & \multicolumn{3}{c}{ $\theta_1$ } &  \multicolumn{3}{c}{$\theta_2$} & \multicolumn{3}{c}{$\theta_3$}\\
    \cline{4-12}
     & & & Bias & SD& RMSE & Bias&SD & RMSE& Bias&SD & RMSE \\
    \hline
  $\Sigma_{ij} = \bm{1}_{\{i=j\}}$ & $n=150$ &  $p=5$ &0.001& 0.021& 0.021& 0.001& 0.029& 0.029& 0.001& 0.026& 0.026  \\
  & &  $p=20$ & 0.001& 0.024& 0.024& 0.001& 0.040& 0.040& 0.001& 0.029& 0.029\\
  & $n=300$ &$p=5$ &0.001& 0.011& 0.011& -0.001& 0.015& 0.015& 0.001& 0.012& 0.012 \\
  &&$p=20$ &0.001&0.011&0.011&0.001&0.017&0.017&0.001&0.013&0.013 \\
  & $n=500$ & $p=5$ &0.001& 0.007& 0.007& -0.001& 0.010& 0.010& 0.001& 0.008& 0.008 \\
  & & $p=20$&0.001& 0.006& 0.006& 0.001& 0.010& 0.010& 0.001& 0.007& 0.007 \\
  \hline
  $\Sigma_{ij} = 0.5^{|i-j|}$ & $n=150$ &  $p=5$ &0.001& 0.024& 0.024& 0.001& 0.032& 0.032& 0.001& 0.026& 0.026 \\
  & & $p=20$ & 0.002& 0.030& 0.030& 0.001& 0.040& 0.040& 0.001& 0.031& 0.031 \\
  & $n=300$ &  $p=5$ & 0.001& 0.011& 0.011& 0.001& 0.016& 0.016& 0.001& 0.012& 0.012\\
  & &$p=20$ & -0.001&0.011&0.011&-0.001&0.016&0.016&0.001&0.013&0.013 \\
  & $n=500$ &  $p=5$ & 0.001& 0.007& 0.007& 0.001& 0.010& 0.010& 0.001& 0.007& 0.008 \\
  &&$p=20$& 0.001& 0.007& 0.007& 0.001& 0.010& 0.010& 0.001& 0.008& 0.008 \\
  \hline
  $\Sigma_{ij} = 1-0.5\cdot\bm{1}_{\{i\neq j\}}$  & $n=150$ &  $p=5$ &0.003& 0.032& 0.032& -0.003& 0.034& 0.034& -0.002& 0.037& 0.037 \\
  & & $p=20$ & 0.002& 0.035& 0.035& -0.003& 0.041& 0.041& 0.001& 0.039& 0.039 \\
  & $n=300$ &  $p=5$ & 0.001& 0.018& 0.018& 0.001& 0.017& 0.017& -0.001& 0.019& 0.019\\
  & & $p=20$ &-0.001& 0.017& 0.017& -0.001& 0.018& 0.018& 0.001& 0.019& 0.019 \\
  & $n=500$ &  $p=5$ &0.001& 0.01& 0.010& 0.001& 0.010& 0.010& 0.001& 0.010& 0.010 \\
  && $p=20$ &0.001& 0.011& 0.011& 0.001& 0.010& 0.010& 0.001& 0.012& 0.012 \\
  \hline
\end{tabular}}
\end{center}
\end{table}

Table \ref{exp2.a.hat} and Table \ref{exp4.a.hat} summarize the estimation performance of the estimated thresholds $\bm{a}$ for the cases with correct
estimation of $\hat{s} = s$ in Examples $2$ and $4$. In both examples, the estimations are of small bias and mean squared error. In fact we note that the jumps
at the two change points are $\|\bm{\delta}_{1}\|^2 = 4$ and $\|\bm{\delta}_{2} \|^2 = 2$, respectively, under both equal and unequal group size situation. In
general it is easier for our methods to estimate the greater jump. In addition, we report the bias and the SD of estimated change plane parameter $\bm{\theta}$
in Table \ref{exp2.theta.hat} and Table \ref{exp4.theta.hat} for Examples 2 and 4 respectively. From the tables, we can conclude that our estimation performs
very well for estimating the change plane parameters.

\begin{figure}[htb!]
\begin{center}
{\tiny
\centerline{
\includegraphics[scale=0.38]{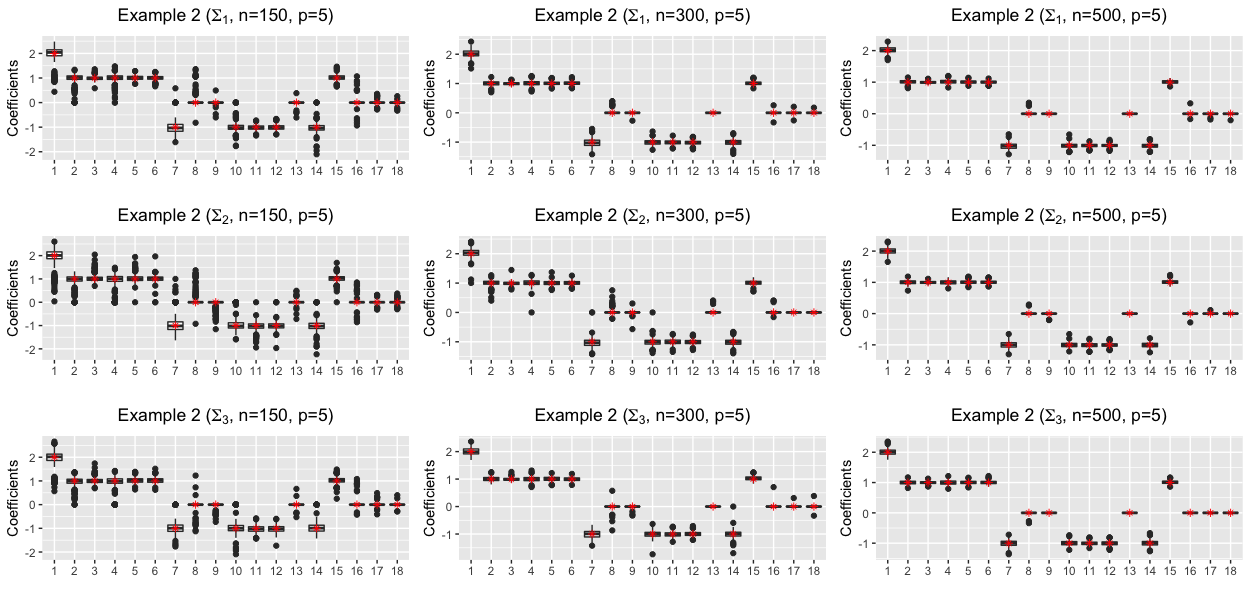}}}
\end{center}
\begin{center}
\vspace{-.4in}
{\tiny
\centerline{
\includegraphics[scale=0.38]{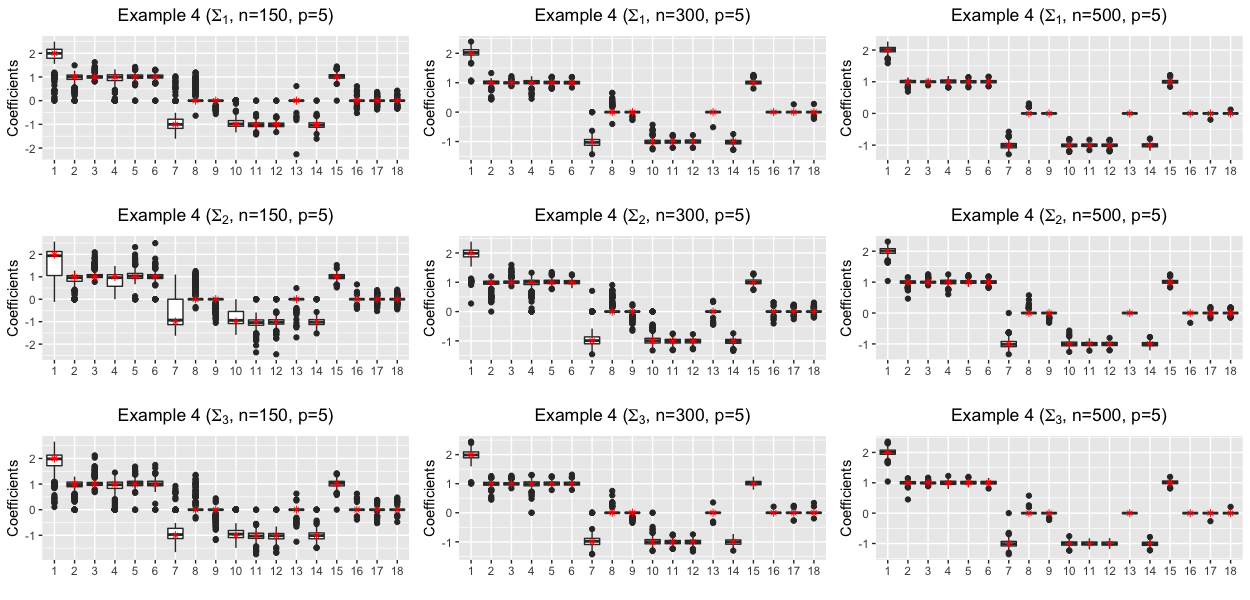}}}
\caption{Box plots of estimated coefficients $\hat{\bm{\gamma}}^{*}$ for Example $2$ and $4$ with three different structures of $\Sigma$.  ``$*$''s are the true
values.}\label{fig:exp24.coef}
\end{center}
\end{figure}

Finally we report the estimation performance of the sparse regression coefficients $\hat{\bm{\gamma}}^{*}$ using boxplots in Figure \ref{fig:exp24.coef}. The
estimated coefficients are all consistent to the true parameter values. The zero coefficients $\gamma_{j}$, $j=8,9,13,16,17,18$, can be accurately identified by
our method. Table \ref{exp24.no0} also shows the number of correctly selected zeros and incorrectly selected zeros in $\hat{\bm{\gamma}}^{*}$, suggesting a
satisfactory variable selection performance.

\begin{table}[h]
\begin{center}
\caption{Variable selection results for Example $2$ and Example $4$ in case of $p=5$. In these cases $6$ coefficients are zero. Correct and Incorrect refer to
the average number of estimated zero coefficients corresponding to the true zero and non-zero coefficients, respectively.}
\label{exp24.no0}
\begin{tabular}{llcccc}
\hline
  && \multicolumn{4}{c}{Avg. no. of $0$ coefficients}\\
  \hline
  && \multicolumn{2}{c}{Example 2} &\multicolumn{2}{c}{Example 4} \\
  \hline
  & &  Correct & Incorrect &Correct & Incorrect \\
  \hline
 $\Sigma_{ij} = \bm{1}_{\{i=j\}}$ & $n=150$ & 5.871& 0.088 &5.649&0.365 \\
  & $n=300$ & 5.971 & 0 & 5.935 & 0.010 \\
  \hline
  $\Sigma_{ij} = 0.5^{|i-j|}$ & $n=150$ & 5.738 & 0.298 & 5.357 & 0.900\\
  & $n=300$ & 5.948 &0.008& 5.658& 0.077 \\
  \hline
  $\Sigma_{ij} = 1-0.5\cdot\bm{1}_{\{i\neq j\}}$ &$n=150$ &5.869 & 0.246&5.706&0.554 \\
  & $n=300$& 5.970 & 0.002& 5.917&0.021 \\
  \hline
\end{tabular}
\end{center}
\end{table}

\section{Application to Real Data}
\label{sec:real}

\subsection{Bovine Collagen Clinical Trial (BCCT)}

We illustrate our methods using clinical data from a 3-year NIH-sponsored randomized Bovine Collagen Trial for Scleroderma patients conducted at 12 centers in
the USA which contains 297 samples (\citealt{postlethwaite2008multicenter, li2009semi}). Patients were randomized to receive oral native collagen at a dose of
500 $ \mu \textnormal{g/day}$ or a placebo. They were evaluated clinically by the Modified Rodnan Skin Score (MRSS) (the primary outcome variable), disability
index of the Health Assessment Questionnaire (HAQ), patient's global assessment, patients pain assessment and physicians global assessment. To implement the
proposed method, we consider $11$ predictor variables: $X_1=$ haq (health assessment questionnaire); $X_2=$ pga (patient self assessment of disease
progression); $X_3=$ dlcop (lung performance measurement 3); $X_4=$ fvcp (lung performance measurement 1); $X_5=$ over (disease progression); $X_6=$ pain (index
of pain); $X_7=$ fev1p (lung performance measurement 2); $X_8 = $ durdis (duration of disease); $X_9 = $ age (in years) $X_{10}=$ ethnic ($0=$ hispanic, $1=$
non-hispanic); $X_{11}=$ sex ($0=$ female, $1 =$ male). Variables are standardized with mean zero and unit variance.

We first fit a linear regression model with $\bm{X}_i = (1, X_{i1}, \ldots, X_{i,11})^{\trans}$ without considering subgroups, and denote
$\hat{\bm{\beta}}^{\textup{ols}}$ the OLS estimation. Then, for subgroup identification, we choose $\bm{Z}_i = (X_{i1}, X_{i2}, X_{i3})^{\trans}$ to be the
threshold variables and fit the multiple threshold change-plane model. The tuning parameters in (\ref{msmobjpen}) were chosen via generalized cross-validation
(GCV). We detect one change by our method with the estimated threshold $\hat{a}^{*} = -0.125$ and the change-plane parameter $\hat{\bm{\theta}}^{*} = (0.801,
-0.206, 0.562)^{\trans}$. The two subgroup sizes are $139$ and $155$ respectively and we report the estimated coefficients $\bm{\beta}$ and $\bm{\delta}$ in
Table \ref{realdata1} with their standard errors (S.E.), and the $p$-values for testing the significance of the coefficients.

\begin{figure}[htb!]
\begin{center}
\includegraphics[scale=0.5]{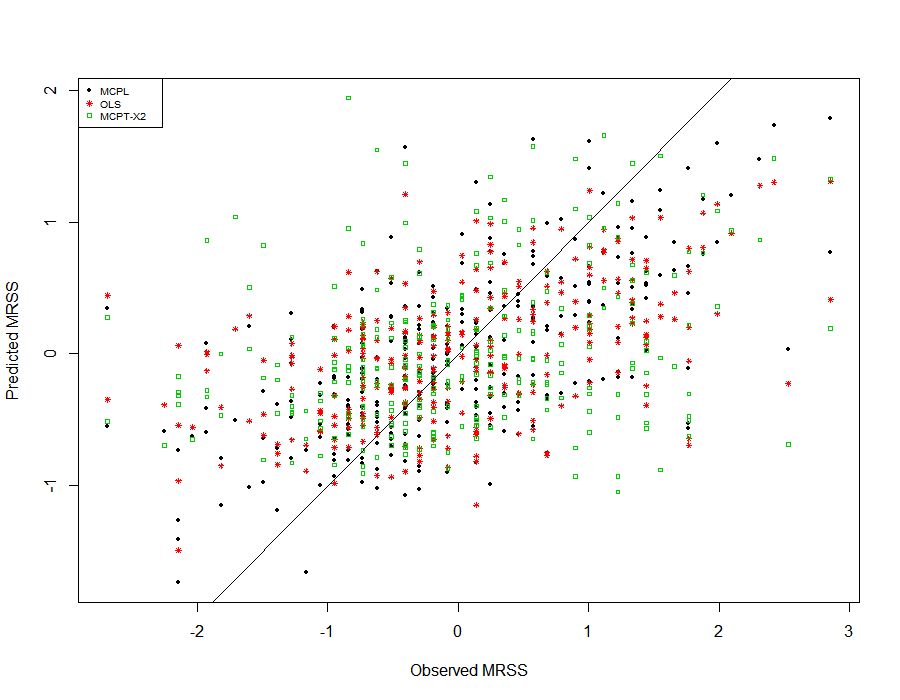}
\caption{The scatter plot of the fitted $Y$ by each method for BCCT data. MCPL stands for multiple threshold change-plane, MCPT-$X_2$,  stands for the MCPT
method with threshold variable $X_2$ and OLS stands for the ordinary least square estimate.}\label{fig:y.realdata1}
\end{center}
\end{figure}

 We compared our MCPL models with the multiple change-points (MCPT) models proposed in \cite{jin2015multi} with single thresholding covariate being $X_1$,
 $X_2$, $X_3$ respectively and also with a version of MCPL with equally weighted plane variable $\bm{Z}_i=(X_{i1}+X_{i2}+X_{i3})/3$ (E-MCPL). From Table
 \ref{realdata1.comp}, we can see that these methods yield quite different subgroups and our proposed MCPL has the smallest mean squared error for predicting
 the MRSS response. In particular, we plot the scatter plots of predicted MRSS versus observed MRSS in Figure \ref{fig:y.realdata1}. One can see that the
 prediction from MCPL is less variable than the other methods.

\begin{figure}[htb]
\centering
  \begin{tabular}{cc}
    \includegraphics[width=.5\textwidth]{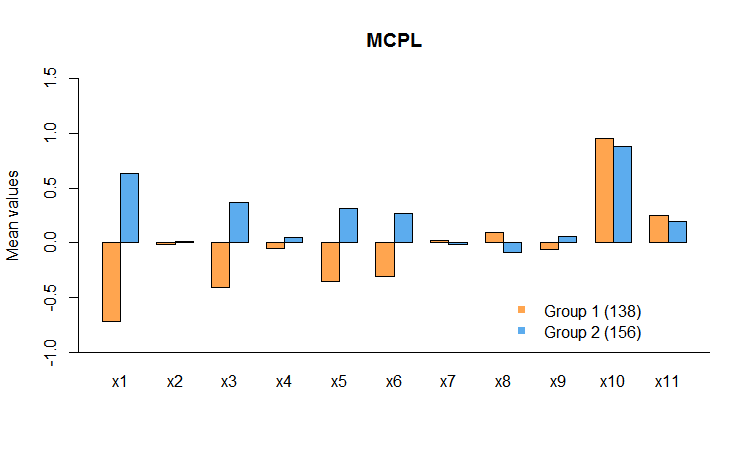}	 &
     \\
    \includegraphics[width=.5\textwidth]{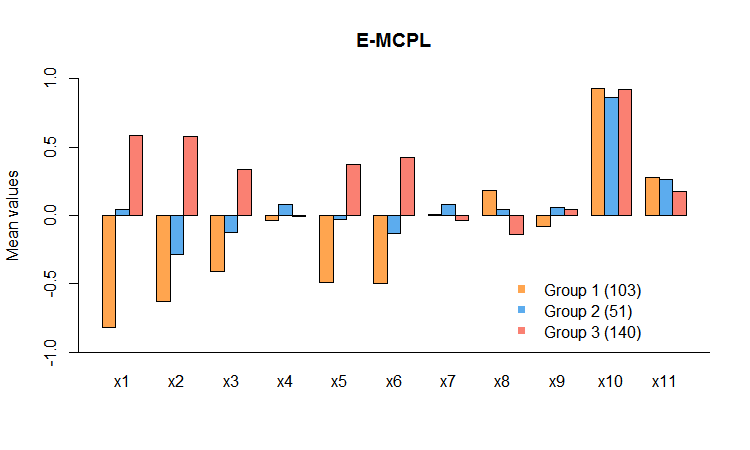} &
    \includegraphics[width=.5\textwidth]{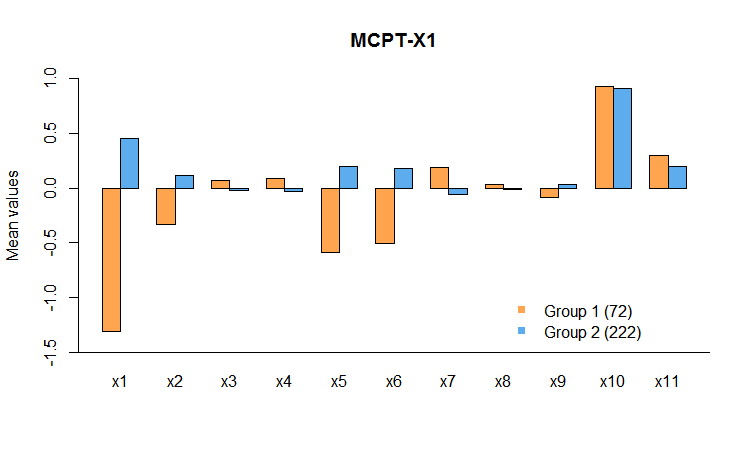} \\
    \includegraphics[width=.5\textwidth]{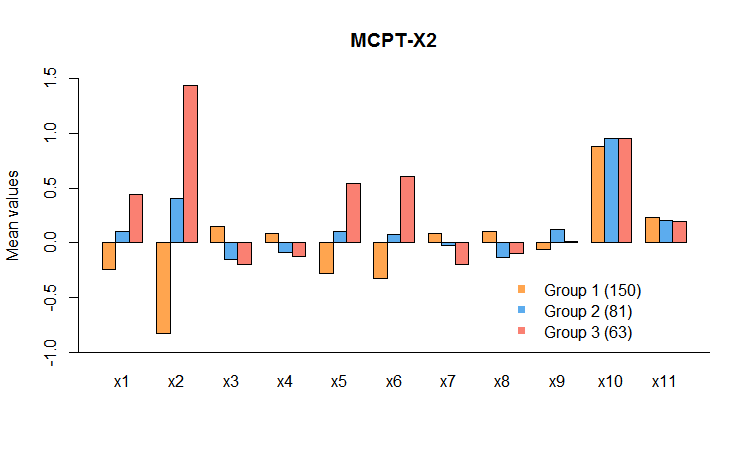} &
    \includegraphics[width=.5\textwidth]{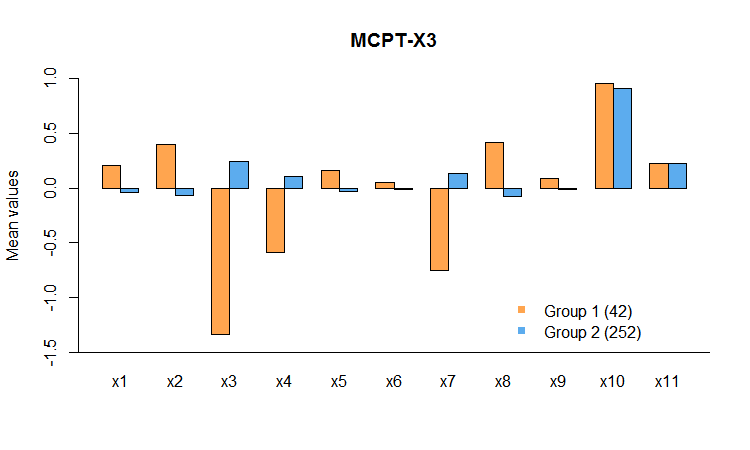}
  \end{tabular}
  \caption{The mean value of predoctors for different detected subgroups for BCCT data. MCPL stands for multiple change-plane model, E-MCPL stands for multiple
  change plane with equal weight, MCPT-$X_1$, MCPT-$X_2$, MCPT-$X_3$ stands for the MCPT method with threshold $X_1$, $X_2$ and $X_3$ respectively. $X_1=$ haq;
  $X_2=$ pga; $X_3=$ dlcop; $X_4=$ fvcp; $X_5=$ over; $X_6=$ pain; $X_7=$ fev1p; $X_8 = $ durdis; $X_9 = $ age  $X_{10}=$ ethnic (non-hispanic); $X_{11}=$ sex
  (male). Group sizes are given in parentheses.}\label{fig:x.group1}
\end{figure}

\begin{figure}[htb]
\centering
  \begin{tabular}{ccc}
    \includegraphics[width=.33\textwidth]{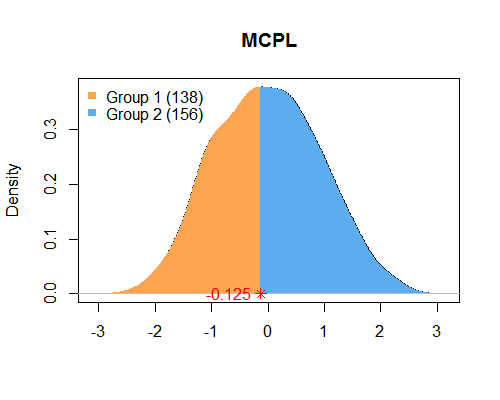} &

    \includegraphics[width=.33\textwidth]{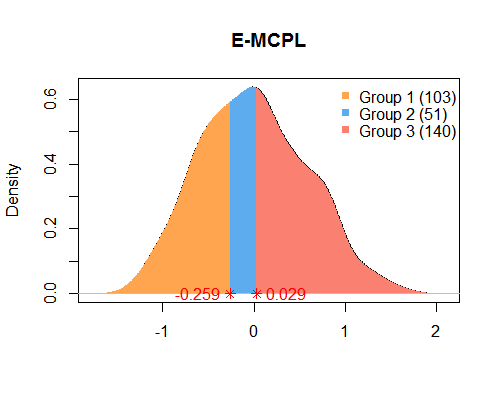}  \\
    \includegraphics[width=.33\textwidth]{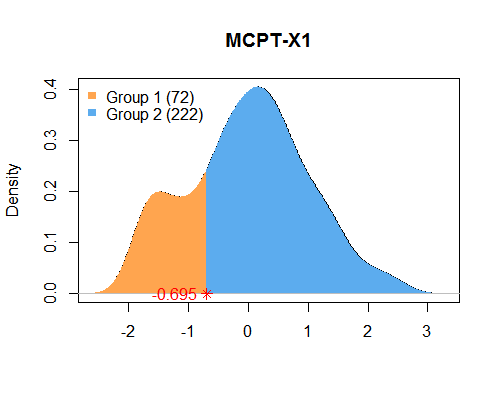} &
    \includegraphics[width=.33\textwidth]{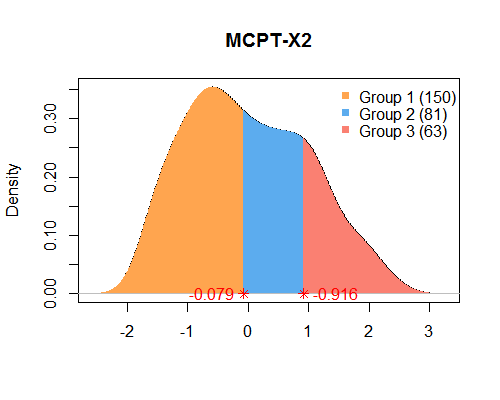} &
    \includegraphics[width=.33\textwidth]{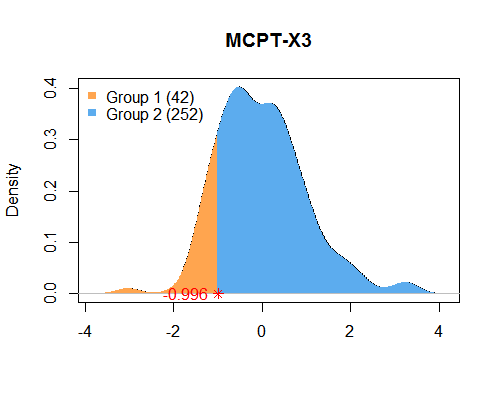}
  \end{tabular}
  \caption{The density plots of thresholding varibles estimated by each method for BCCT data, the red mark points in x-axis disply the cut-off points. MCPL
  stands for  multiple change-plane, SCPL stands for  single change-plane, E-MCPL stands for multiple change plane with equal weight, MCPT-$X_1$, MCPT-$X_2$,
  MCPT-$X_3$ stands for the MCPT method with threshold $X_1$, $X_2$ and $X_3$ respectively. Group sizes are given in parentheses.}\label{fig:w.dens1}
\end{figure}

 To gain more understanding of the groups, we summarize the means of all covariates for the detected subgroups in Figure \ref{fig:x.group1}. Eyeballing the
 plots we can see that the mean difference of all the covariates between the two subgroups detected by MCPL are quite different compared to the other methods.
 We also plot the kernel density estimation of the thresholding variable ${\bf Z}^T\hat{\bm\theta}$ for all methods in Figure \ref{fig:w.dens1}.

\begin{table}[h]\footnotesize
\begin{center}
\caption{Estimated results for Bovine Collagen Clinical Trial (BCCT), along with standard errors (S.E.) and P-values. $X_1=$ haq; $X_2=$ pga; $X_3=$ dlcop;
$X_4=$ fvcp; $X_5=$ over; $X_6=$ pain; $X_7=$ fev1p; $X_8 = $ durdis; $X_9 = $ age  $X_{10}=$ ethnic (non-hispanic); $X_{11}=$ sex (male).}
\label{realdata1}
\begin{tabular}{lcccccccccc}
\hline
  & \multicolumn{3}{c}{ $\bm{\beta}$ } &  \multicolumn{3}{c}{$\bm{\delta}$} & \multicolumn{3}{c}{ $\bm{\beta}^{\textup{ols}}$ }\\
    \cline{2-10}
 Covariates  & Coef. & S.E. & P-value & Coef. & S.E. & P-value & Coef. & S.E. & P-value \\
    \hline
   \textbf{Intercept}& 0.022&0.104 &0.830 &-0.525& 0.238 & 0.028 &-0.383&0.188&0.042  \\
   $X_1$ & 0 & - & - & 0.443& 0.125 & $<0.001$&0.220&0.065&$<0.001$ \\
   $X_2$ & 0.416 &0.057&$<0.001$ & 0& - & - &0.369&0.059& $<0.001$\\
   $X_3$ & 0.332 & 0.102 & 0.001 & -0.261& 0.131& 0.048 &0.113&0.058&0.051\\
   $X_4$ & 0.124 & 0.081 & 0.127 & 0& - & - &0.118&0.086&0.170\\
   $X_5$  & -0.296 & 0.128 & 0.022 & 0.547 & 0.173 & 0.002 &0.065&0.092&0.480\\
   $X_6$ & 0.245 & 0.122 & 0.045  &-0.602 & 0.160 & $<0.001$ &-0.095&0.087&0.276\\
   $X_7$ & -0.322 & 0.099 & 0.001 &0.307& 0.111 & 0.006 &-0.121&0.084&0.153 \\
   $X_8$ & -0.187 & 0.068 & 0.006  &0.313& 0.105 & 0.003  &-0.063&0.055&0.251\\
   $X_9$ & -0.104 & 0.050 & 0.037 & 0& - & - &-0.115&0.053&0.031\\
   $X_{10}$ & 0 & - & -& 0.367 & 0.209 & 0.080 &0.386&0.195&0.049\\
   $X_{11}$ & -0.376 & 0.172 & 0.029 & 0.771&0.247& 0.002 &0.138&0.130&0.287\\
  \hline
  %MSE & \multicolumn{6}{c}{ 0.644} & \multicolumn{3}{c}{0.739} \\
  %\hline
\end{tabular}
\end{center}
\end{table}

\begin{table}[h]
\begin{center}
\caption{Estimated Comparison for Bovine Collagen Clinical Trial (BCCT). MCPL stands for  multiple change-plane,  MCPT-$X_1$, MCPT-$X_2$, MCPT-$X_3$ stands for
the MCPT method with threshold $X_1$, $X_2$ and $X_3$ respectively, E-MCPL stands for multiple change plane with equal weight and OLS stands for the ordinary
least square estimate.}
\label{realdata1.comp}
\begin{tabular}{lcccc}
\hline
 Model  & MSE & $\hat{s}$ & threshod $\hat{a}$ & Group Sizes   \\
    \hline
MCPL & 0.613 & 1& $-0.125$ & $138:156$ \\
MCPT-$X_1$& 0.695 & 1 & $-0.695$ & $72 : 222$\\
MCPT-$X_2$& 0.672 & 2 &  $(-0.079, 0.916)$ & $150:81:63$\\
MCPT-$X_3$ & 0.786 & 1 & $-0.996 $ & $42 : 252$\\
E-MCPL &0.707 & 2 & $(-0.259, 0.029)$ & $103:51:140$ \\
%Change-plane & 0.739 & 0 & - &- \\
OLS & 0.739 & 0& -& -\\
  \hline
\end{tabular}
\end{center}
\end{table}

\subsection{AIDS Clinical Trials}

We apply our method to the AIDS Clinical Trials Group Study 175 (ACTG175), which contains 2139 HIV-infected subjects. This randomized clinical trial compares
zidovudine (ZDV) monotherapy (treatment 0) with other three therapies including ZDV and didanosine (ddI) (treatment 1), ZDV and zalcitabine (zal) (treatment 2),
and ddI monotherapy (treatment 3) in adults infected with the human immunodeficiency virus type I  (\citealt{tsiatis2008covariate, lu2013variable}). Our
interest is to conduct subgroup analysis to produce more satisfactory predicted value of CD4 counts (cells/mm$^3$) at $20\pm 5$ weeks. We consider the following
covariates: $X_1=$ hemophilia (0 =no, 1 =yes); $X_2=$ gender (0 =female, 1 =male); $X_3=$ CD4 counts at baseline; $X_4=$ antiretroviral history (0 =naive, 1
=experienced); $X_5=$ age (years); $X_6=$ weight (kg); $X_7=$ Karnofsky score; $X_8=$ CD8 counts at baseline; $X_{9}=$ homosexual activity (0 =no, 1 =yes);
$X_{10}=$ history of intravenous drug use (0 =no, 1 =yes); $X_{11}=$ race (0 =white, 1 =white);  $X_{12}=$ symptomatic status (0 =asymptomatic, 1 =symptomatic)
and $X_{13}=$ treatment arm (0=zidovudine, 1=zidovudine and didanosine, 2=zidovudine and zalcitabine, 3=didanosine).

We first fit a linear regression model with $\bm{X}_i = (1, X_{i1}, \ldots, X_{i,14})^{\trans}$ without subgroups, and denote $\hat{\bm{\beta}}^{\textup{ols}}$
the OLS estimation. We then fit the MCPL model (\ref{mcp}), and choose $\bm{Z}_i = (X_{i1}, X_{i2}, X_{i3}, X_{i4}, X_{i5})^{\trans}$ as the threshold
variables. Similarly, the tuning parameters in (\ref{msmobjpen}) were chosen via the GCV. The estimated change-plane parameter $\hat{\bm{\theta}}^{*} = (-0.268,
-0.199, 0.876, -0.268, 0.223)^{\trans}$. We detect two change planes by our method where the estimated threshold locations are $\hat{\bm{a}}^{*} = (-0.309,
0.201)^{\trans}$, thus producing three subgroups with group sizes $1162$, $394$, and $583$ respectively.  Table \ref{realdata2.all} reports the estimated
coefficients $\bm{\beta}$ and $\bm{\delta}$, their standard errors (S.E.), and the $p$-values for testing the significance of the coefficients.

\begin{figure}[htb!]
\begin{center}
\includegraphics[scale=0.5]{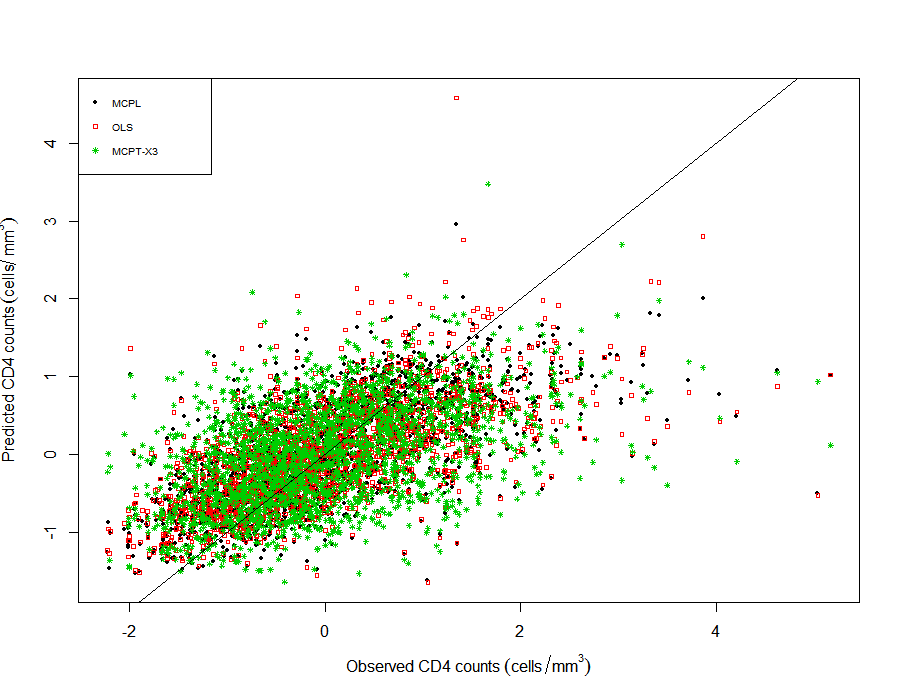}
\caption{The scatter plot of the fitted $Y$ by each method for ACTG 175 data. MCPL stands for multiple change-plane, MCPT-$X_3$,  stands for the MCPT method
with threshold  $X_3$ and OLS stands for the ordinary least square estimate.}\label{fig:y.realdata2}
\end{center}
\end{figure}

\begin{figure}[htb]
\centering
  \begin{tabular}{c}
    \includegraphics[width=.55\textwidth]{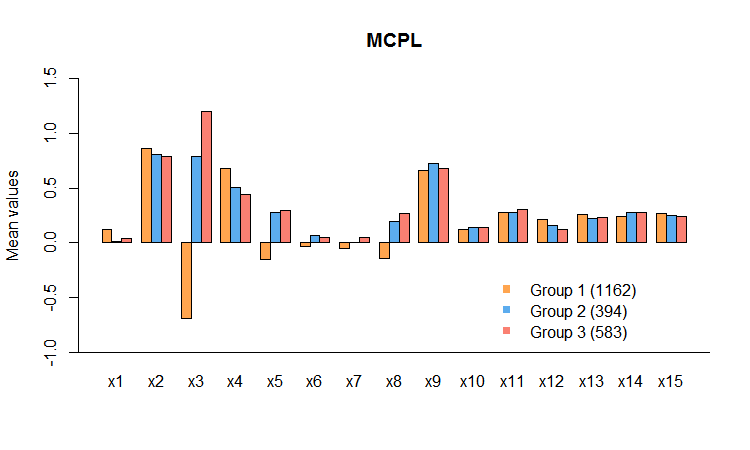} \\
    \includegraphics[width=.55\textwidth]{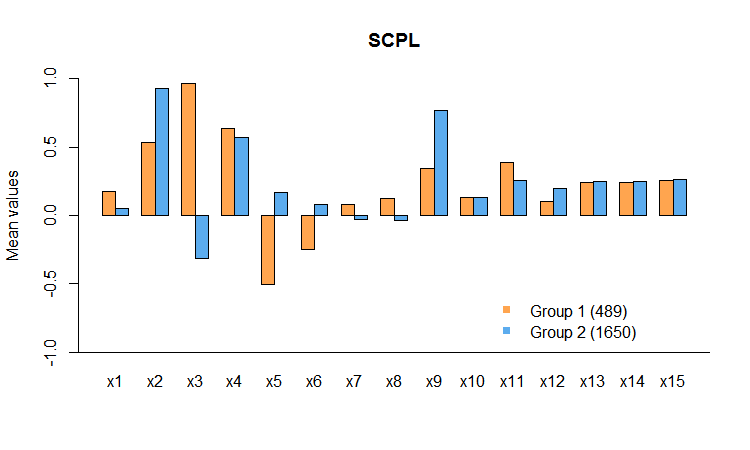} \\
    \includegraphics[width=.55\textwidth]{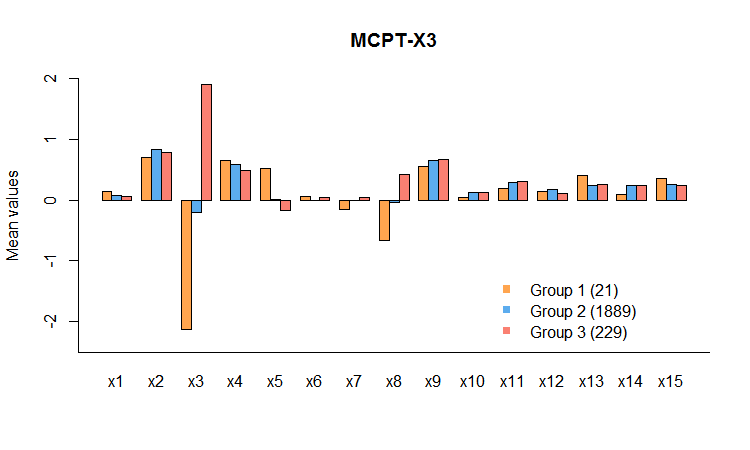}
  \end{tabular}
  \caption{The mean value of predoctors for different detected subgroups for ACTG 175 data. MCPL stands for multiple change-plane, SCPL stands for single
  change-plane, MCPT-$X_3$ stands for the MCPT method with threshold $X_3$. $X_1=$ hemophilia (yes); $X_2=$ gender (male); $X_3=$ CD4 counts at baseline; $X_4=$
  antiretroviral history (experienced); $X_5=$ age; $X_6=$ weight; $X_7=$ Karnofsky score; $X_8=$ CD8 counts at baseline; $X_{9}=$ homosexual activity (yes);
  $X_{10}=$ history of intravenous drug use (yes); $X_{11}=$ race (white);  $X_{12}=$ symptomatic status (symptomatic), $X_{13}=$ treatment arm$^1$ (zidovudine
  and didanosine), $X_{14}=$ treatment arm$^2$ (zidovudine and zalcitabine) and $X_{15}=$ treatment arm$^3$ (didanosine). Group sizes are given in
  parentheses.}\label{fig:x.group2}
\end{figure}

\begin{figure}[htb]
\centering
  \begin{tabular}{ccc}
    \includegraphics[width=.3\textwidth]{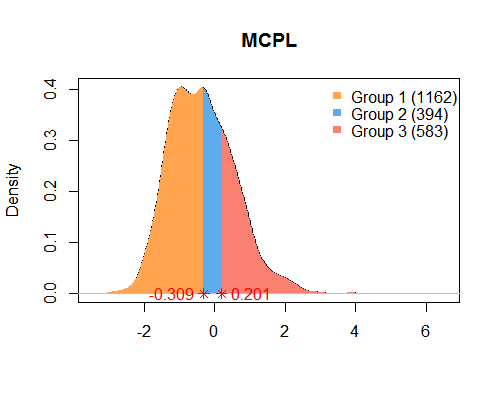} &
    \includegraphics[width=.3\textwidth]{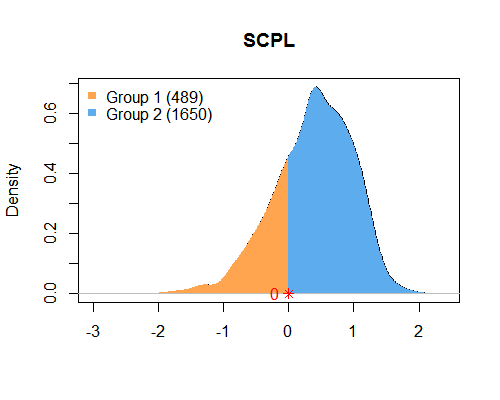} &
    \includegraphics[width=.3\textwidth]{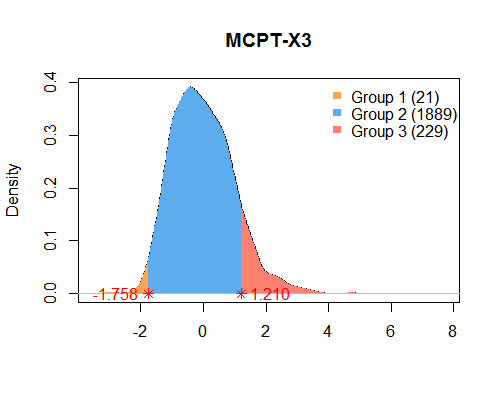}
  \end{tabular}
  \caption{The density plots of thresholding varibles estimated by each method for ACTG 175 data, the red mark points in x-axis disply the cut-off points. MCPL
  stands for multiple change-plane, SCPL stands for single change-plane, MCPT-$X_3$ stands for the MCPT method with threshold $X_3$. Group sizes are given in
  parentheses.}\label{fig:w.dens2}
\end{figure}

We also compared the prediction performance of our MCPL models with the single-threshold change-plane (SCPL) models (\ref{scp}), with the multiple change-points
(MCPT) models proposed in \cite{jin2015multi} with single threshold covariate being $X_3$ and $X_5$ respectively and also with a version of MCPL with equally
weighted plane multiple variables $\bm{Z}_i=(X_1+\cdots+X_5)/5$ (E-MCPL). In this case, $X_1$, $X_2$ and $X_4$ are not continuous and cannot be applied in MCPT
model. The MSE results from all these methods are summarized in table \ref{realdata2.all.comp} and we can see that MCPL achieves the smallest MSE. Furthermore,
we display the scatter plots of predicted CD4 counts versus observed CD4 counts in Figure \ref{fig:y.realdata2}. We can draw similar conclusion as in the first
example. To study the subgroups, we summarize the means of all the covariates for the subgroup in Figure \ref{fig:x.group2}. We also plot the kernel density
plots of the thresholding variables for all methods in Figure \ref{fig:w.dens2}.

\begin{table}[h]\footnotesize
\begin{center}
\caption{Estimated results for AIDS Clinical Trials Group Study 175 (ZDV vs.\ the other three treatments), along with standard errors (S.E.) and P-values by t
test. $X_0=$ Intercept; $X_1=$ hemophilia (yes); $X_2=$ gender (male); $X_3=$ CD4 counts at baseline; $X_4=$ antiretroviral history (experienced); $X_5=$ age; $X_6=$ weight;
$X_7=$ Karnofsky score; $X_8=$ CD8 counts at baseline; $X_{9}=$ homosexual activity (yes); $X_{10}=$ history of intravenous drug use (yes); $X_{11}=$ race
(white);  $X_{12}=$ symptomatic status (symptomatic), $X_{13}=$ treatment arm$^1$ (zidovudine and didanosine), $X_{14}=$ treatment arm$^2$ (zidovudine and
zalcitabine) and $X_{15}=$ treatment arm$^3$ (didanosine).}
\label{realdata2.all}
\begin{tabular}{lcccccccccccc}
\hline
  & \multicolumn{3}{c}{ $\bm{\beta}$ } &  \multicolumn{3}{c}{$\bm{\delta}_1$} & \multicolumn{3}{c}{$\bm{\delta}_2$}& \multicolumn{3}{c}{
  $\bm{\beta}^{\textup{ols}}$ }\\
    \cline{2-13}
   & Coef. & S.E. & P-value & Coef. & S.E. & P-value & Coef. & S.E. & P-value& Coef. & S.E. & P-value \\
    \hline
  $X_0$ &-0.087&0.059 &0.143 &0.171&0.078 &0.029 &0.286&0.092 &0.002 &-0.001&0.061&0.981\\
  $X_1$ &-0.169&0.062&0.007 &0 &-&-&0&-&-&-0.184&0.080&0.021 \\
  $X_2$ &0&-&-&0&-&-&0&-&-&-0.032&0.066&0.632 \\
  $X_3$ &0.568&0.041 &$<0.001$ &0&-&- &-0.269&0.058&$<0.001$&0.571&0.017&$<0.001$ \\
  $X_4$ &-0.247&0.035 &$<0.001$ &0&-&-&0&-&-&-0.281&0.035&$<0.001$\\
  $X_5$ &0&-&- &-0.117&0.025&$<0.001$ &0&- & -&0.021&0.017&0.234 \\
  $X_6$ &0 & - &- &0&-&-&0&- &- &-0.003&0.018&0.846 \\
  $X_7$ &0.085&0.021&  $<0.001$ &-0.233&0.047 & $<0.001$ &0.187& 0.055& 0.001&0.040&0.017&0.019 \\
  $X_8$ &-0.109 &0.024 &$<0.001$&0.085&0.034&0.012&0&-&-&-0.066&0.017&$<0.001$ \\
  $X_9$ &0&-&- &-0.064&0.057&0.257&0&-&-&0.003&0.059&0.957 \\
  $X_{10}$ &0.092&0.058&0.111&0&- &- & -0.167&0.109 &0.128 &0.052&0.052&0.319\\
  $X_{11}$ &-0.101&0.043&0.020&0&- &-&-0.128&0.083&0.123&-0.128&0.040&0.002 \\
  $X_{12}$ &-0.135&0.055 &0.015 &0.103&0.120&0.390&-0.110&0.145&0.450  &-0.130&0.045&0.004\\
  $X_{13}$ &0.452&0.061&$<0.001$&0.164&0.107&0.125 &-0.124&0.121&0.309&0.488&0.048&$<0.001$ \\
  $X_{14}$  & 0.251&0.062&$<0.001$&0.008&0.086&0.927&0&-&-&0.252&0.048&$<0.001$ \\
  $X_{15}$ &0.228&0.052&$<0.001$&0&-&-&0.261&0.096&0.007&0.294&0.047&$<0.001$ \\
  \hline
  %MSE & \multicolumn{9}{c}{0.568} &  \multicolumn{3}{c}{0.595} \\
  %\hline
\end{tabular}
\end{center}
\end{table}

\begin{table}[h]
\begin{center}
\caption{Estimated Comparison for AIDS Clinical Trials Group Study 175 data (ZDV vs.\ the other three treatments). MCPL stands for multiple change-plane, SCPL
stands for single change-plane, MCPT-$X_3$, MCPT-$X_5$stands for the MCPT method with threshold $X_3$ and $X_5$ respectively, E-MCPL stands for multiple change
plane with equal weight and OLS stands for the ordinary least square estimate.}
\label{realdata2.all.comp}
\begin{tabular}{lcccc}
\hline
 Method  & MSE & $\hat{s}$ & Threshold $\hat{a}$ & Group Sizes \\
    \hline
MCPL & 0.567 & 2& $(-0.309, 0.201)$ & $1162:394:583$ \\
SCPL & 0.578& 1& 0& $489:1650$ \\
MCPT-$X_3$ & 0.575 & 2 & $(-1.758, 1.210)$ & $21:1889:229$\\
MCPT-$X_5$ & 0.595& 0&-&-\\
E-MCPL & 0.595 & 0& -& - \\
%Change-plane & 0.595 & 0 & - &- \\
OLS & 0.595 & 0& -& - \\
  \hline
\end{tabular}
\end{center}
\end{table}

\section{Discussion}
\label{sec:dis}

In our theoretical results, we allow the coefficients of the covariates to be sparse, but require their dimension to be much smaller than $n$. A high or
ultra-high dimensional situation can be further investigated (\citealt{shi2017high}). Our proposed method can be extended to other models including generalized
linear models and hazard regression models to incorporate non-Gaussian response variables. Although these extensions appear to be conceptually straightforward,
it is a nontrivial task to develop computational algorithms and establish theoretical properties in these more complicated models.

%\bigskip
%\begin{center}
%{\large\bf Appendix: Proofs of Theorems}
%\end{center}

\section*{Supplementary Materials}
The supplementary materials contain technical proofs for Theorems 1-3.

%\clearpage
%\appendix

\clearpage
\bibliographystyle{chicago}
\bibliography{changeplane}

\end{document}